\documentclass[journal]{vgtc}                

\usepackage{mathptmx}
\usepackage{graphicx}
\usepackage{times}
\usepackage{subcaption}
\usepackage{amsmath}
\usepackage{epstopdf}
\usepackage{amsfonts}
\usepackage{csquotes}
\usepackage{balance}
\usepackage{paralist}

\usepackage{gensymb}

\usepackage[noend]{algpseudocode}
\usepackage{algorithmicx,algorithm}

\usepackage[bookmarks,backref=true,linkcolor=black]{hyperref} 
\hypersetup{
	pdfauthor = {},
	pdftitle = {},
	pdfsubject = {},
	pdfkeywords = {},
	colorlinks=true,
	linkcolor= black,
	citecolor= black,
	pageanchor=true,
	urlcolor = black,
	plainpages = false,
	linktocpage
}

\onlineid{1096}

\vgtccategory{Research}

\title{
	Uncertainty-Oriented Ensemble Data Visualization and Exploration using Variable Spatial Spreading
}

\author{Mingdong Zhang, Li Chen, Quan Li, Xiaoru Yuan, \textit{Senior Member, IEEE}, and Junhai Yong}
\authorfooter{
	\item
	Mingdong Zhang, Li Chen, and Junhai Yong are with School of Software, BNRist, Tsinghua University, Beijing, China. E-mail: zhangmd14@mails.tsinghua.edu.cn; {chenlee,yongjh}@tsinghua.edu.cn.
	\item
	Quan Li is with AI Group, WeBank, Shenzhen, Guangdong, China. E-mail: forrestli@webank.com.
	\item 
	Xiaoru Yuan is with Key Laboratory of Machine Perception (Ministry of Education), and National Engineering Laboratory for Big Data Analysis and Application, Peking University, Beijing, China. E-mail: xiaoru.yuan@pku.edu.cn.
	\item 
	Li Chen is the corresponding author.
}

\shortauthortitle{Correll and Heer: Surpise! Bayesian Weighting of Thematic Maps}

\abstract{As an important method of handling potential uncertainties in numerical simulations, ensemble simulation has been widely applied in many disciplines. Visualization is a promising and powerful ensemble simulation analysis method. However, conventional visualization methods mainly aim at data simplification and highlighting important information based on domain expertise instead of providing a flexible data exploration and intervention mechanism. Trial-and-error procedures have to be repeatedly conducted by such approaches. To resolve this issue, we propose a new perspective of ensemble data analysis using the attribute variable dimension as the primary analysis dimension. Particularly, we propose a variable uncertainty calculation method based on variable spatial spreading. Based on this method, we design an interactive ensemble analysis framework that provides a flexible interactive exploration of the ensemble data. Particularly, the proposed spreading curve view, the region stability heat map view, and the temporal analysis view, together with the commonly used 2D map view, jointly support uncertainty distribution perception, region selection, and temporal analysis, as well as other analysis requirements.  We verify our approach by analyzing a real-world ensemble simulation dataset. Feedback collected from domain experts confirms the efficacy of our framework.} 


\newcommand{\frameTeaser}{	
	\teaser{
		\centering		
		\includegraphics[width=.9\textwidth]{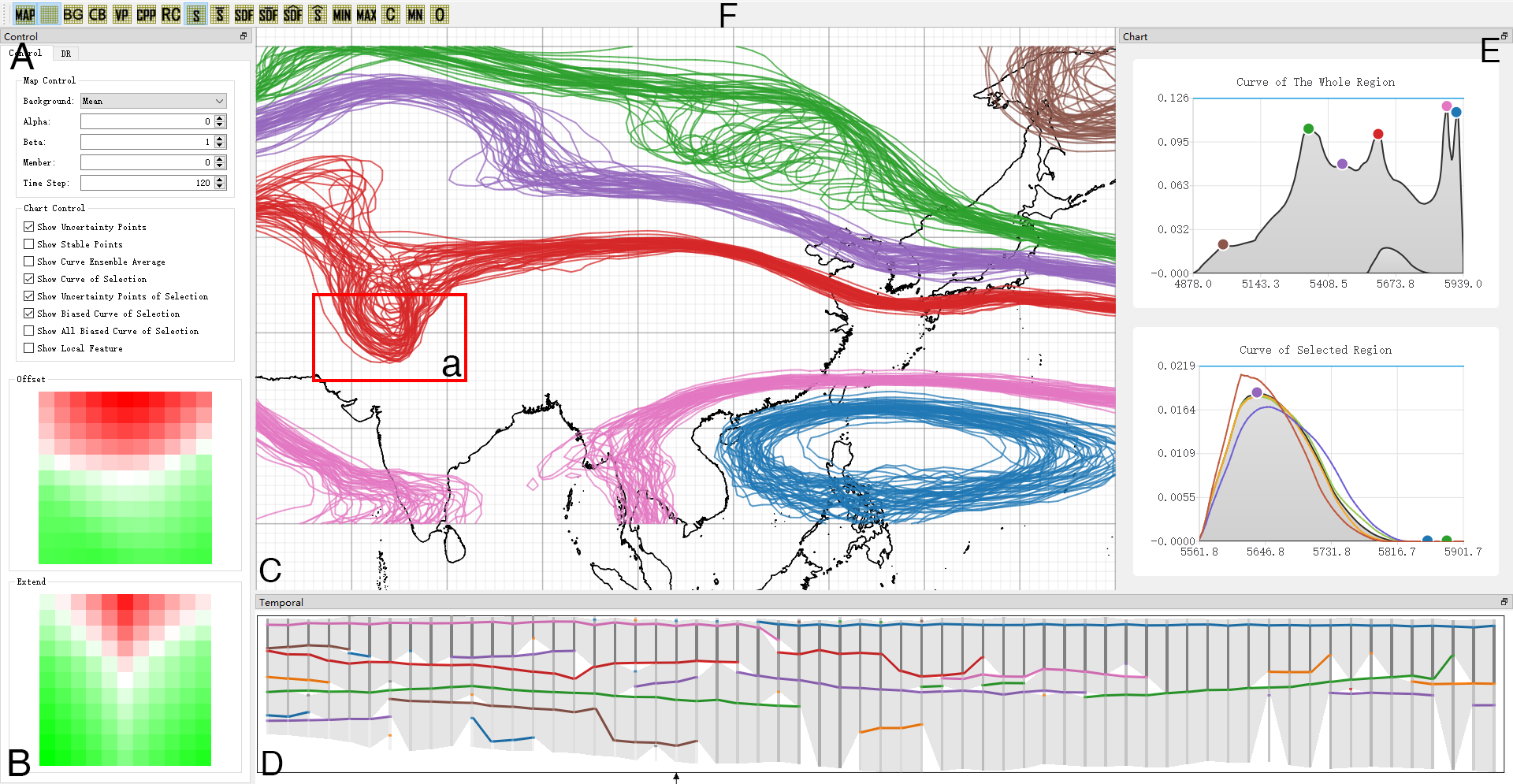}	
			\vspace{-2mm}	
		\caption{Uncertainty-oriented ensemble data visualization framework interface: (A) The parameter setting panel provides control of the visualization parameters. (B) The region stability heat map view shows the stability of the selected region and provides region adjustment through direct clicking. (C) The 2D map view shows the features of the selected isovalues and integrates up-to-date visualization methods. (D) The temporal analysis view shows the temporal relationships of the features and supports temporal selection. (E) The spatial spreading curve view shows the spatial spreading of the variable bins globally (top) and locally (bottom). 
		(F) The display control toolbar enables switching between different visualization methods.	 	
		}
		\label{fig:frame}
	}	
}

\newcommand{\meanspreadFig}{	
	\begin{figure*}[h]
		\centering
		\vspace{-2mm}
		\begin{subfigure}[tb]{0.32\textwidth}
			\includegraphics[width=\linewidth]{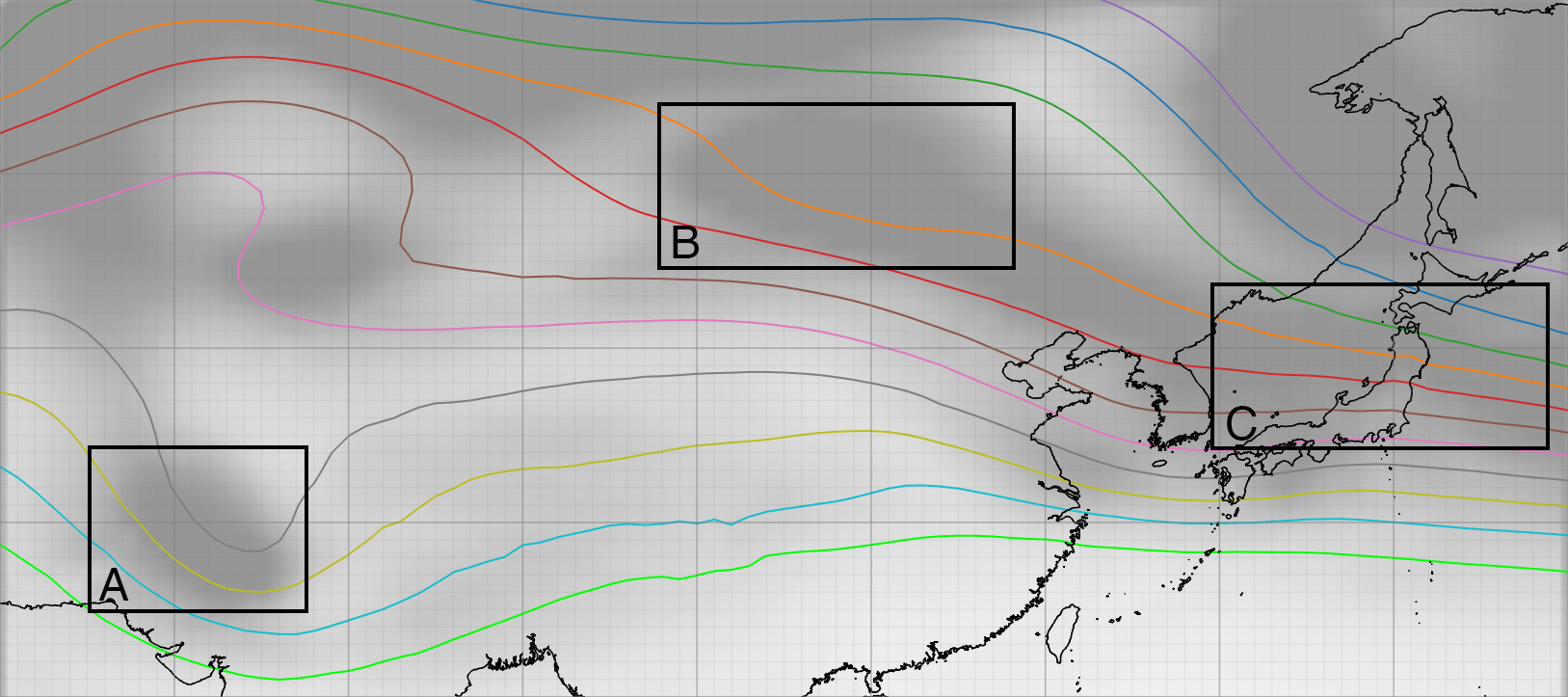}
			\caption{Contours of ensemble mean}
			\label{fig:meanspread_mean}
		\end{subfigure}
		~	
		\begin{subfigure}[tb]{0.32\textwidth}
			\includegraphics[width=\linewidth]{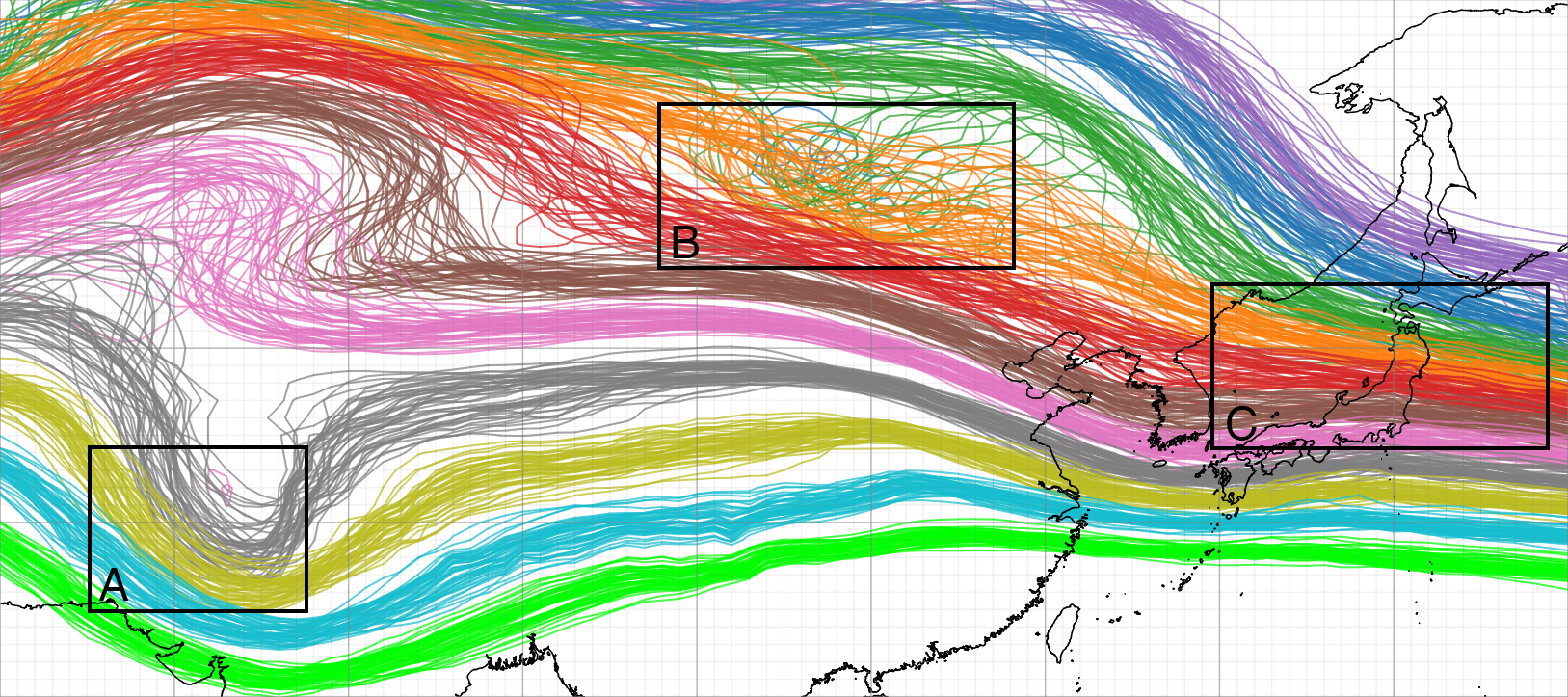}
			\caption{Spaghetti plots}
			\label{fig:meanspread_s}
		\end{subfigure}
		~	
		\begin{subfigure}[tb]{0.32\textwidth}
			\includegraphics[width=\linewidth]{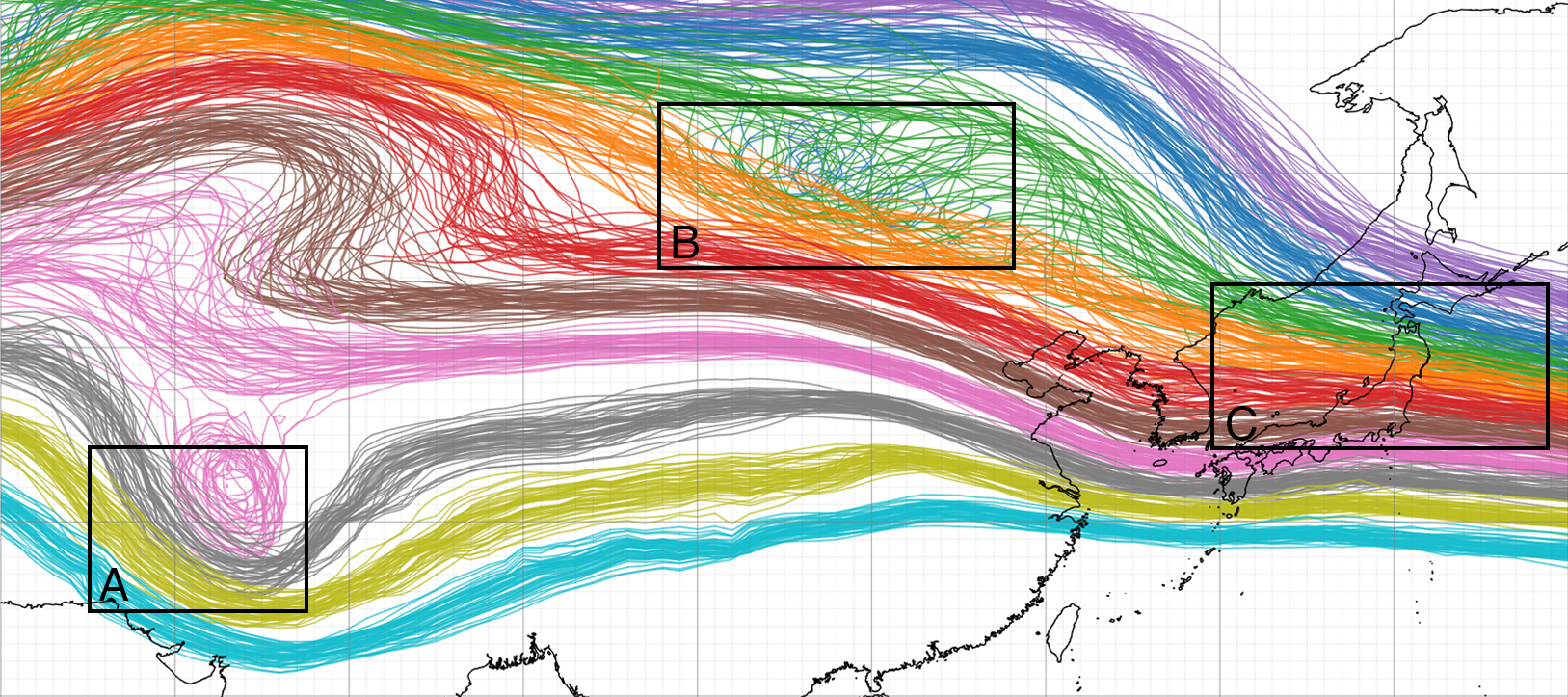}
			\caption{Spaghetti plots}
			\label{fig:meanspread_s_2}
		\end{subfigure}
		\vspace{-2mm}
		\caption{Contours of ensemble mean and (b, c) spaghetti plots are drawn for the ensemble forecast data of geopotential height. In (a), the variance of the ensemble is mapped to the background color.  The isovalues of (a) and (b) are \(5200,5260,...,5800\) (kpm), and \(5230, 5290,...,5830\) (kpm) of (c), respectively. Three regions with high uncertainty are selected. The reasons for this selection are analyzed in Section~\ref{sec:requirement}.}
		\label{fig:meanspread}
		\vspace{-4mm}
	\end{figure*}	
}

\newcommand{\mappingFig}{	
	\begin{figure}[h]
		\centering
		\vspace{-2mm}
		\begin{subfigure}[tb]{0.48\linewidth}
			\includegraphics[width=\linewidth]{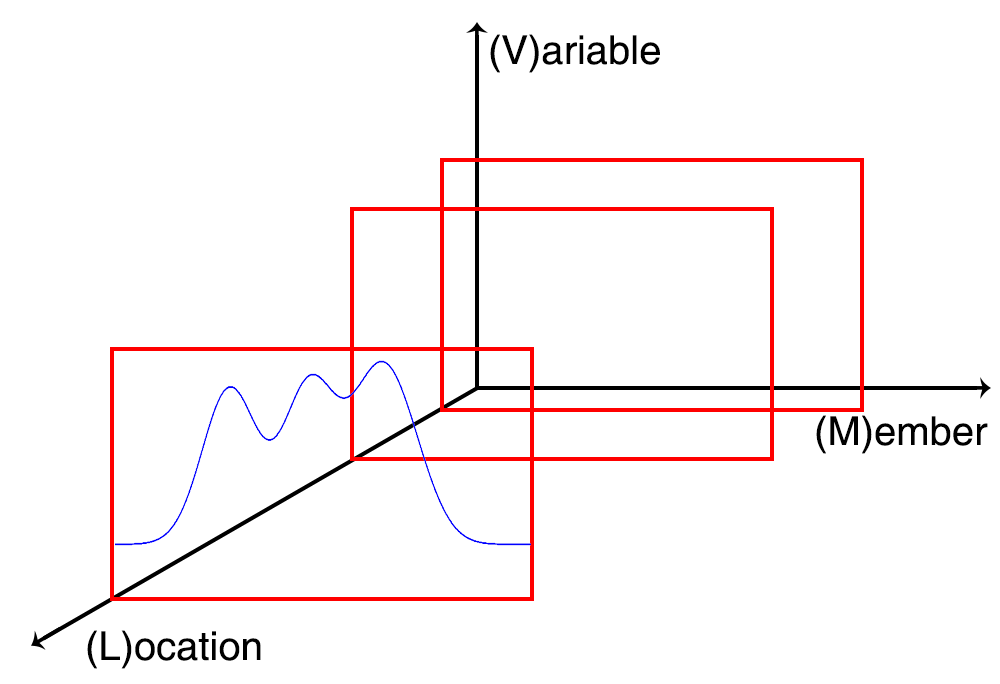}
			\caption{Location oriented}
			\label{fig:mapping_location}
		\end{subfigure}	
		~
		\begin{subfigure}[tb]{0.48\linewidth}
			\includegraphics[width=\linewidth]{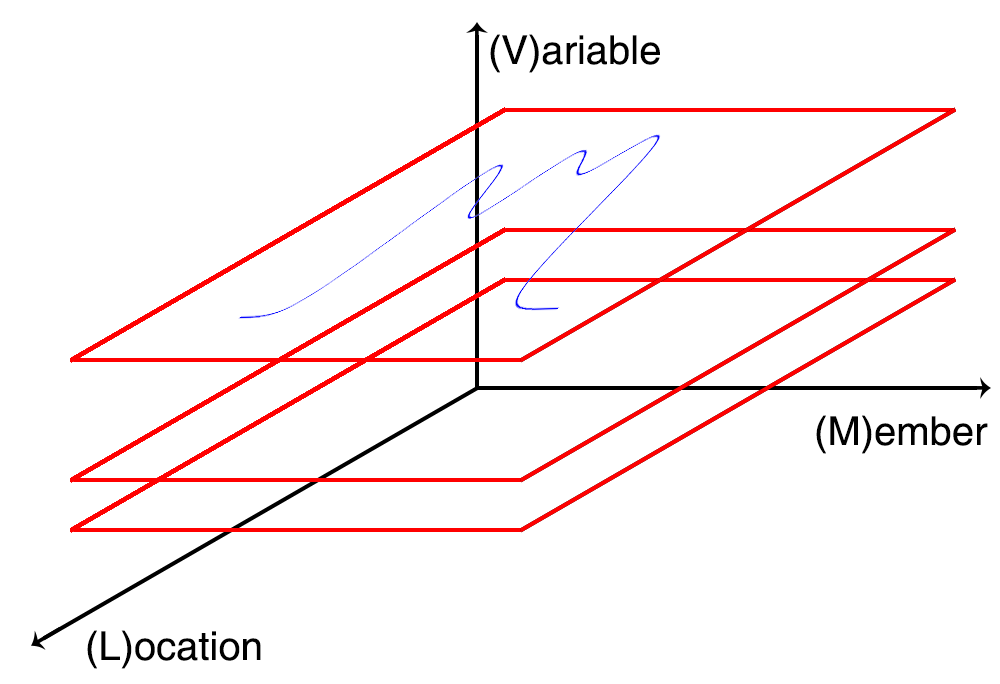}
			\caption{Variable oriented}
			\label{fig:mapping_variable}
		\end{subfigure}
		\vspace{-2mm}
		
		\caption{Analysis using (a) location as the primary dimension versus (b) using variable as the primary dimension.}
		\label{fig:mapping}
	\end{figure}	
}

\newcommand{\standDeviationFig}{	
	\begin{figure*}[h]
		\centering
		\vspace{-2mm}
		\begin{subfigure}[tb]{.3\linewidth}
			\includegraphics[width=\linewidth]{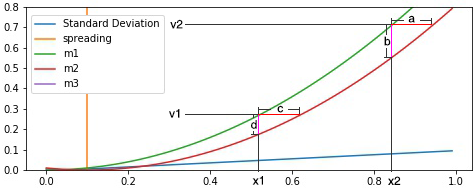}
			\caption{}
			\label{fig:spreading12}
		\end{subfigure}	
		~
		\begin{subfigure}[tb]{.3\linewidth}
			\includegraphics[width=\linewidth]{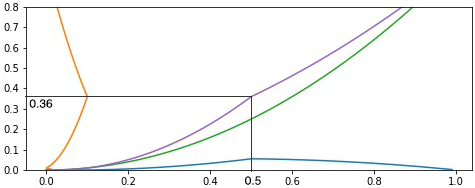}
			\caption{}
			\label{fig:spreading13}
		\end{subfigure}	
		~
		\begin{subfigure}[tb]{.3\linewidth}
			\includegraphics[width=\linewidth]{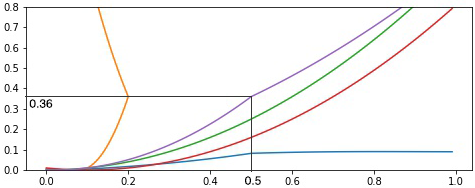}
			\caption{}
			\label{fig:spreading123}
		\end{subfigure}	
		\vspace{-2mm}
		\caption{Demonstration of variable spreading. (a) Case 1: Equally distributed uncertainty in ensemble of two members can be reflected by variable spreading but they are inconsistent with standard deviation. (b) Case 2: Variable spreading and standard deviation can both reflect the maximum point of uncertainty in the ensemble with two members. (c) Case 3: Variable spreading can show the maximum point of uncertainty when a third member is added, but the standard deviation cannot.}
		\label{fig:spreading}
		\vspace{-4mm}
	\end{figure*}	
}

\newcommand{\parameterAFig}{	
	\begin{figure}[h]
		\centering
		\vspace{-2mm}

		\begin{subfigure}[h]{.31\linewidth}
			\includegraphics[width=\linewidth]{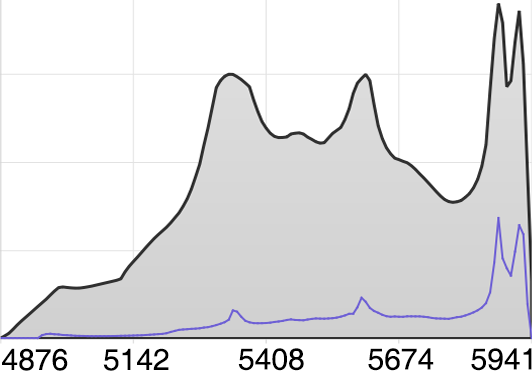}
			\caption{\(\alpha=0\)}
			\label{fig:parameterA_0}
		\end{subfigure}	
		~
		\begin{subfigure}[h]{.31\linewidth}
			\includegraphics[width=\linewidth]{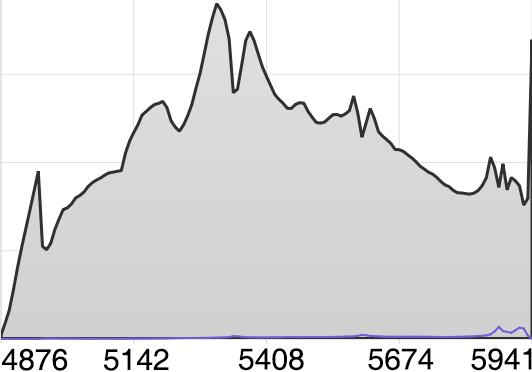}
			\caption{\(\alpha=0.5\)}
			\label{fig:parameterA_5}
		\end{subfigure}	
		~
		\begin{subfigure}[h]{.31\linewidth}
			\includegraphics[width=\linewidth]{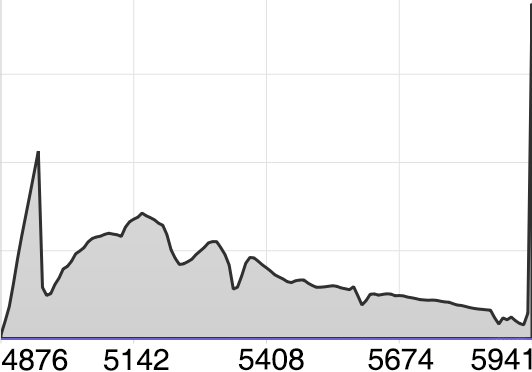}
			\caption{\(\alpha=1\)}
			\label{fig:parameterA_1}
		\end{subfigure}			
		\vspace{-2mm}
		\caption{Spatial spreading curves of different \(\alpha\). The black curve with gray filling is the calculated result, whereas the blue curves are the corresponding curves of the average field.}
		\label{fig:parameterA}
	\end{figure}	
}

\newcommand{\regionComparisonFig}{	
	\begin{figure}[h]
			\vspace{-2mm}
		\centering		
		\begin{subfigure}[tb]{.361\linewidth}
			\includegraphics[width=\linewidth]{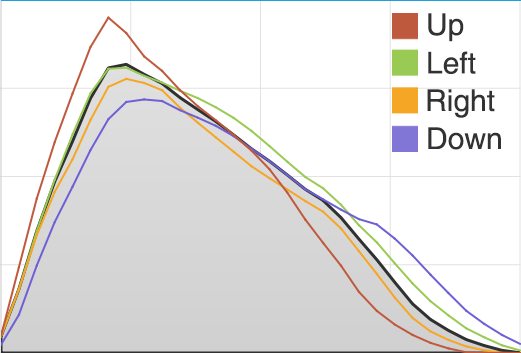}
			\caption{Biased spreading curve}
			\label{fig:region_comparison_curve_1}
		\end{subfigure}	
		~
		\begin{subfigure}[tb]{.326\linewidth}
			\includegraphics[width=\linewidth]{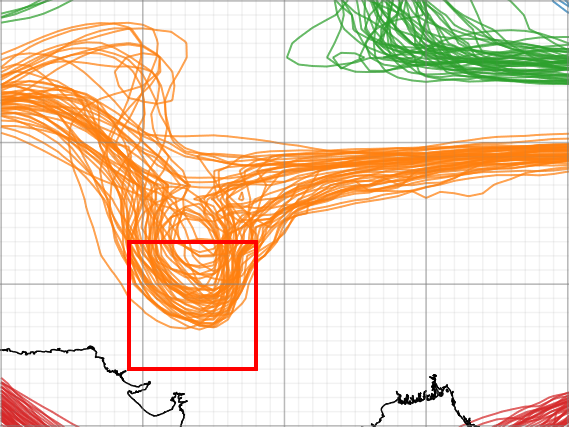}
			\caption{Selected region}
			\label{fig:region_comparison_spaghetti_1}
		\end{subfigure}	
		~
		\begin{subfigure}[tb]{.263\linewidth}
			\includegraphics[width=\linewidth]{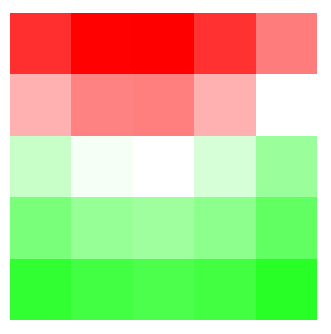}
			\caption{Stability heat map}
			\label{fig:region_comparison_heatmap_1}
		\end{subfigure}			
		
		\begin{subfigure}[tb]{.361\linewidth}
			\includegraphics[width=\linewidth]{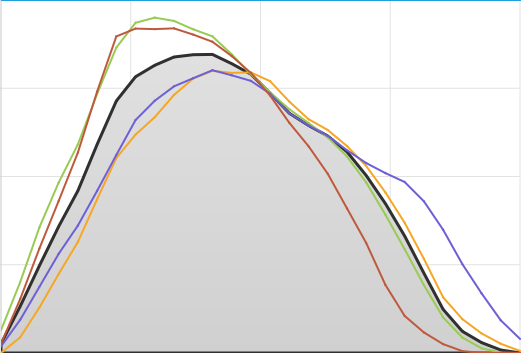}
			\caption{Biased spreading curve}
			\label{fig:region_comparison_curve_2}
		\end{subfigure}	
		~
		\begin{subfigure}[tb]{.326\linewidth}
			\includegraphics[width=\linewidth]{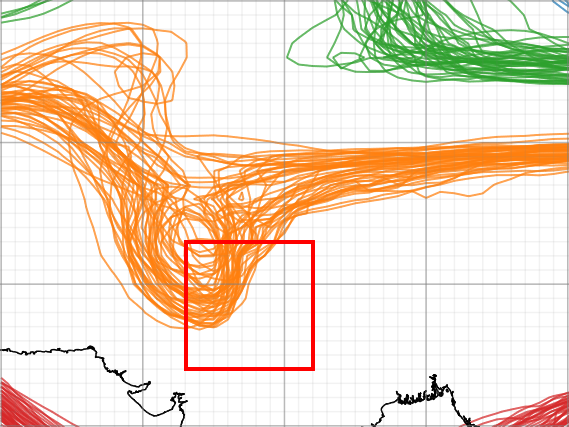}
			\caption{Selected region}
			\label{fig:region_comparison_spaghetti_2}
		\end{subfigure}	
		~
		\begin{subfigure}[tb]{.263\linewidth}
			\includegraphics[width=\linewidth]{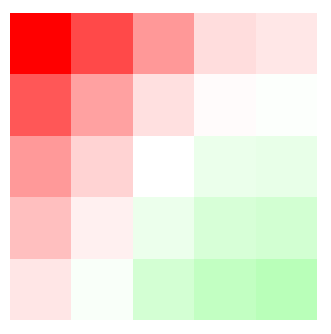}
			\caption{Stability heat map}
			\label{fig:region_comparison_heatmap_2}
		\end{subfigure}			
		
		\begin{subfigure}[tb]{.361\linewidth}
			\includegraphics[width=\linewidth]{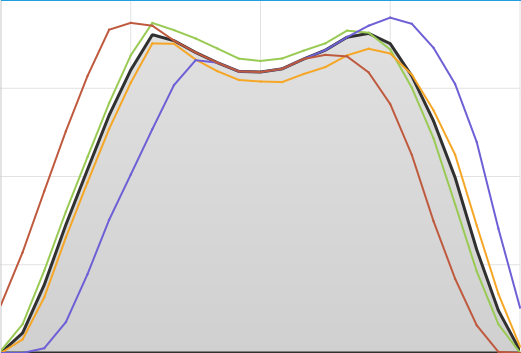}
			\caption{Biased spreading curve}
			\label{fig:region_comparison_curve_3}
		\end{subfigure}	
		~
		\begin{subfigure}[tb]{.326\linewidth}
			\includegraphics[width=\linewidth]{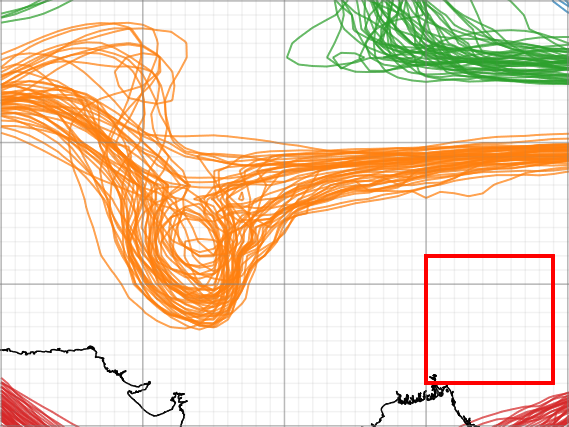}
			\caption{Selected region}
			\label{fig:region_comparison_spaghetti_3}
		\end{subfigure}	
		~
		\begin{subfigure}[tb]{.263\linewidth}
			\includegraphics[width=\linewidth]{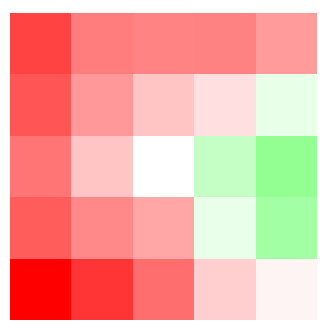}
			\caption{Stability heat map}
			\label{fig:region_comparison_heatmap_3}
		\end{subfigure}		
	
		\caption{Offset line charts and heat maps of different region selections. The middle column shows the region selections with the spaghetti plot of the globally selected maximum points as the background. The first and third columns show the corresponding offset line chart and stability heat map, respectively.}
			\vspace{-2mm}
		\label{fig:region_comparison}
	\end{figure}	
}

\newcommand{\criticalPointFig}{	
	\begin{figure}[h]
		\centering		
		
		\begin{subfigure}[tb]{.48\linewidth}
			\includegraphics[width=\linewidth]{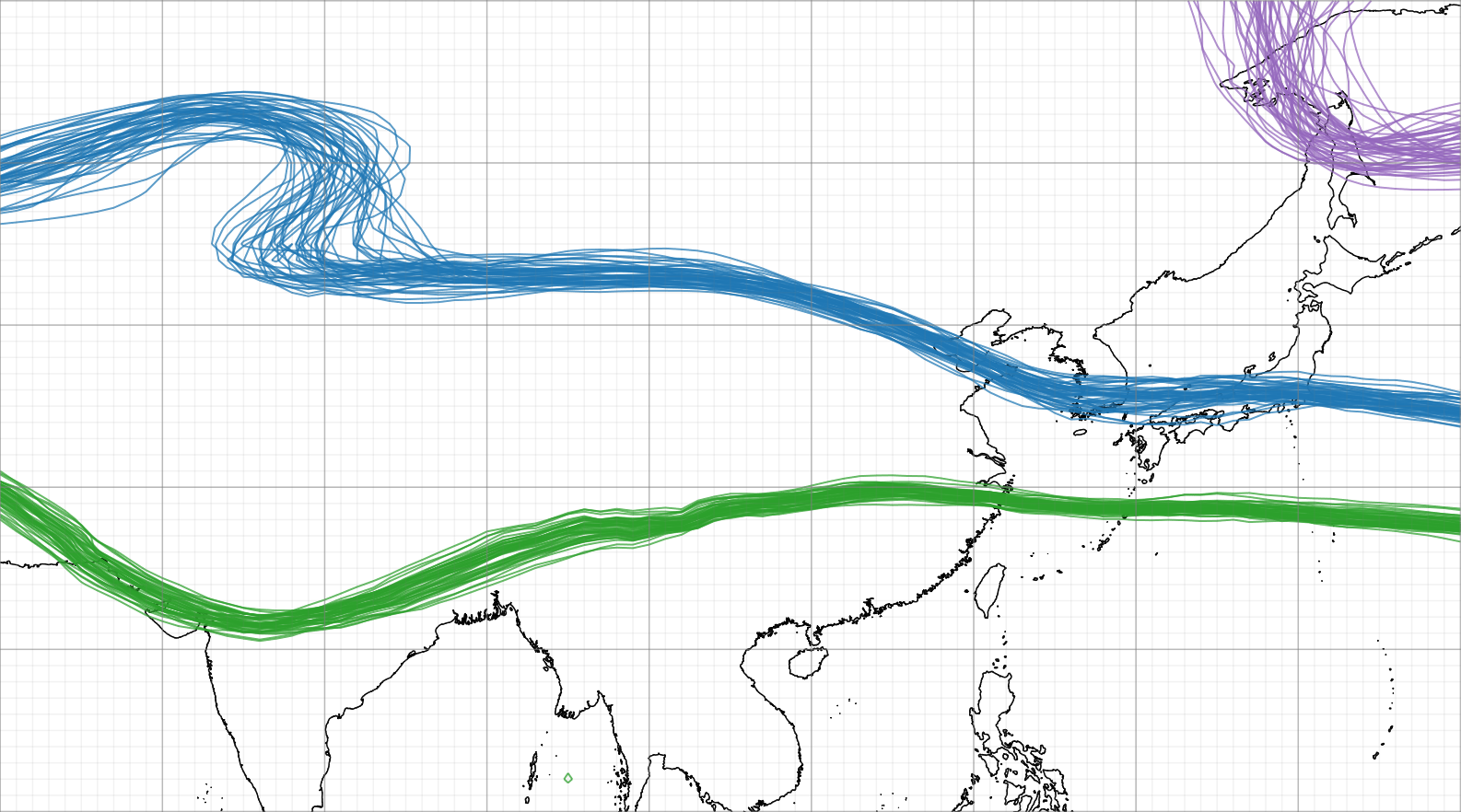}
			\caption{SPs}
			\label{fig:stable_points}
		\end{subfigure}	
		~
		\begin{subfigure}[tb]{.48\linewidth}
			\includegraphics[width=\linewidth]{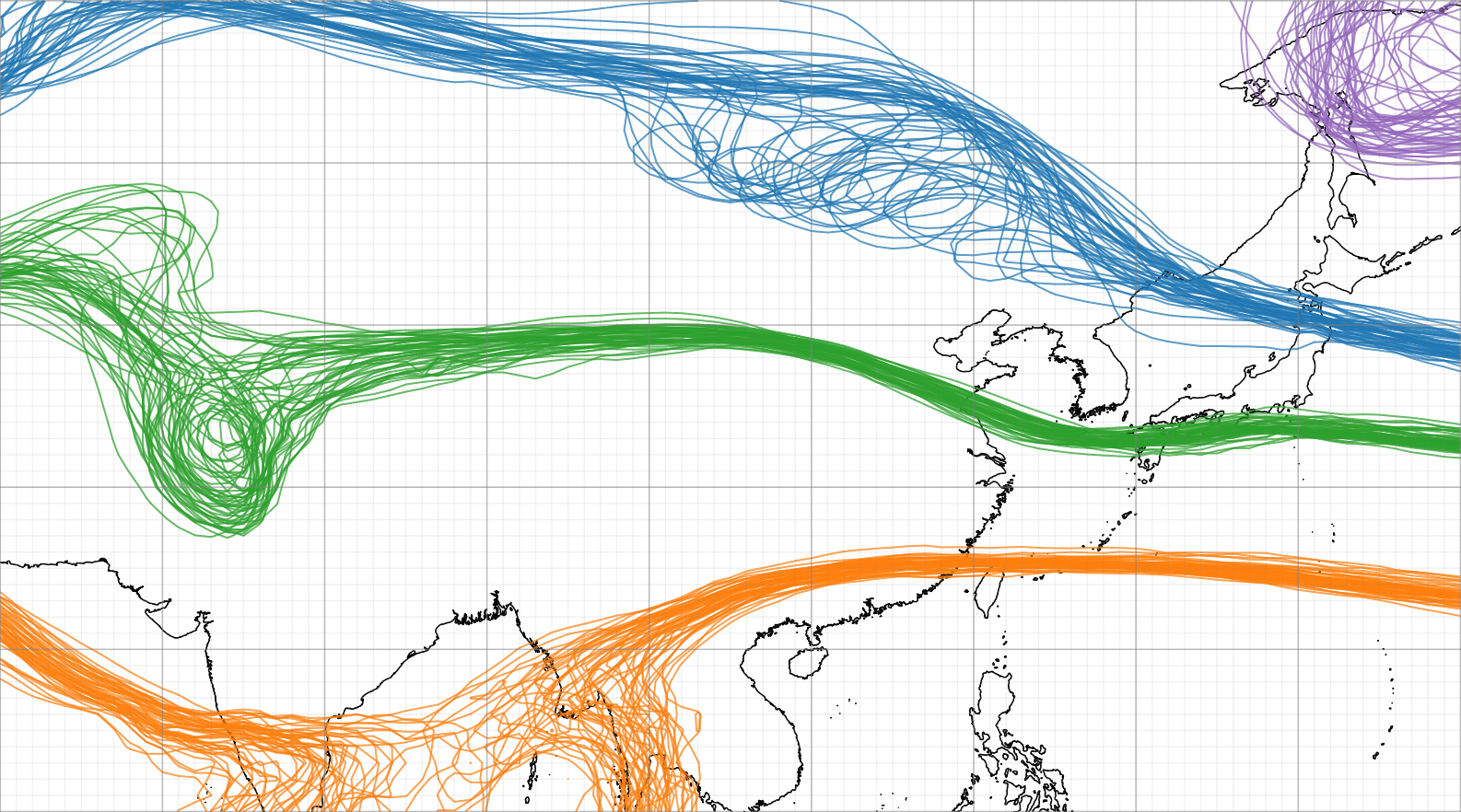}
			\caption{UPs}
			\label{fig:uncertainty_points}
		\end{subfigure}	
		
		\vspace{-2mm}
		\caption{Spaghetti plots of extreme points on the spreading curve.}
			\vspace{-2mm}
		\label{fig:critical_points}
	\end{figure}	
}

\newcommand{\samplingPointsFig}{	
	\begin{figure}[h]
		\centering
		\vspace{-2mm}
		\begin{subfigure}[h]{.48\linewidth}
			\includegraphics[width=\linewidth]{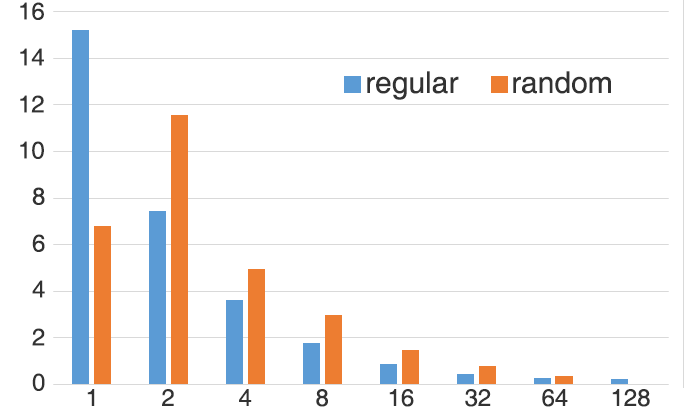}
			\caption{}
			\label{fig:sampling_points_randombase}
		\end{subfigure}	
		~
		\begin{subfigure}[h]{.48\linewidth}
			\includegraphics[width=\linewidth]{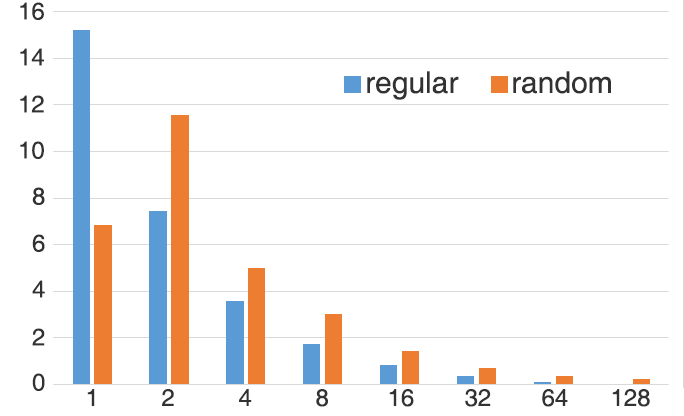}
			\caption{}
			\label{fig:sampling_points_regularbase}
		\end{subfigure}			
		\vspace{-2mm}
		\caption{Comparison of the effect of the different numbers of sampling points of regular and random sampling. For fairness, we use a sampling number of 128-times grid points for  (a) random and (b) regular sampling as the baseline.}
		\vspace{-2mm}
		\label{fig:sampling_points}
	\end{figure}	
}

\newcommand{\globalFig}{	
	\begin{figure}[h]
		\centering
		
		\begin{subfigure}[tb]{\linewidth}
			\includegraphics[width=\linewidth]{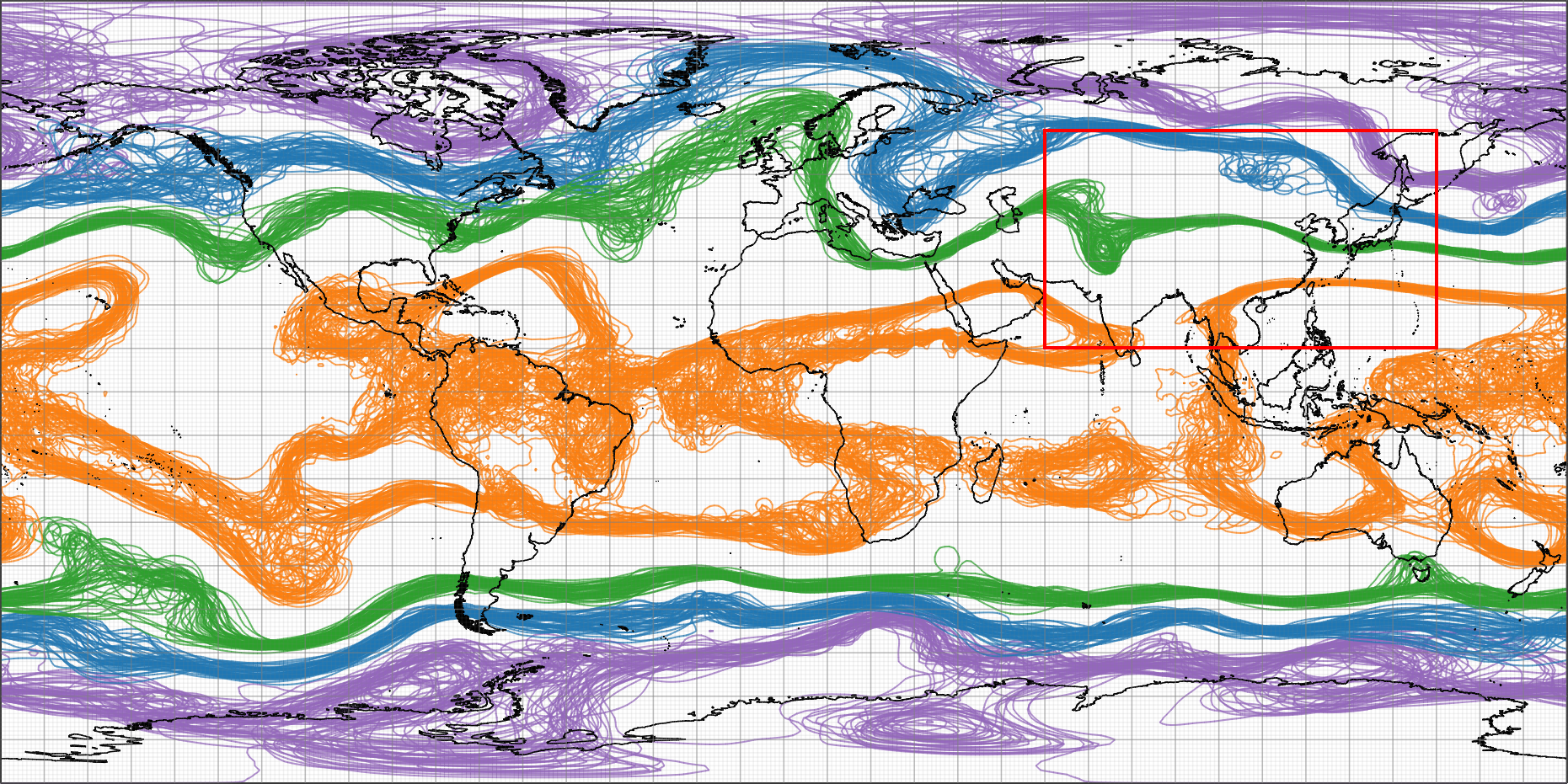}
			\caption{Spaghetti plot of the global region.}
			\label{fig:spaghetti_global}
		\end{subfigure}	
		
		\begin{subfigure}[tb]{0.455\linewidth}
			\includegraphics[width=\linewidth]{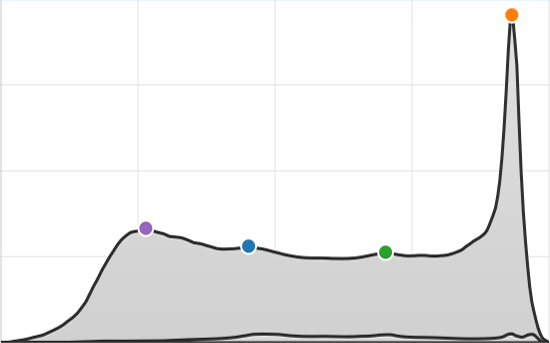}
			\caption{Spreading curve calculated globally}
			\label{fig:curve_global}
		\end{subfigure}
		~
		\begin{subfigure}[tb]{0.51\linewidth}
			\includegraphics[width=\linewidth]{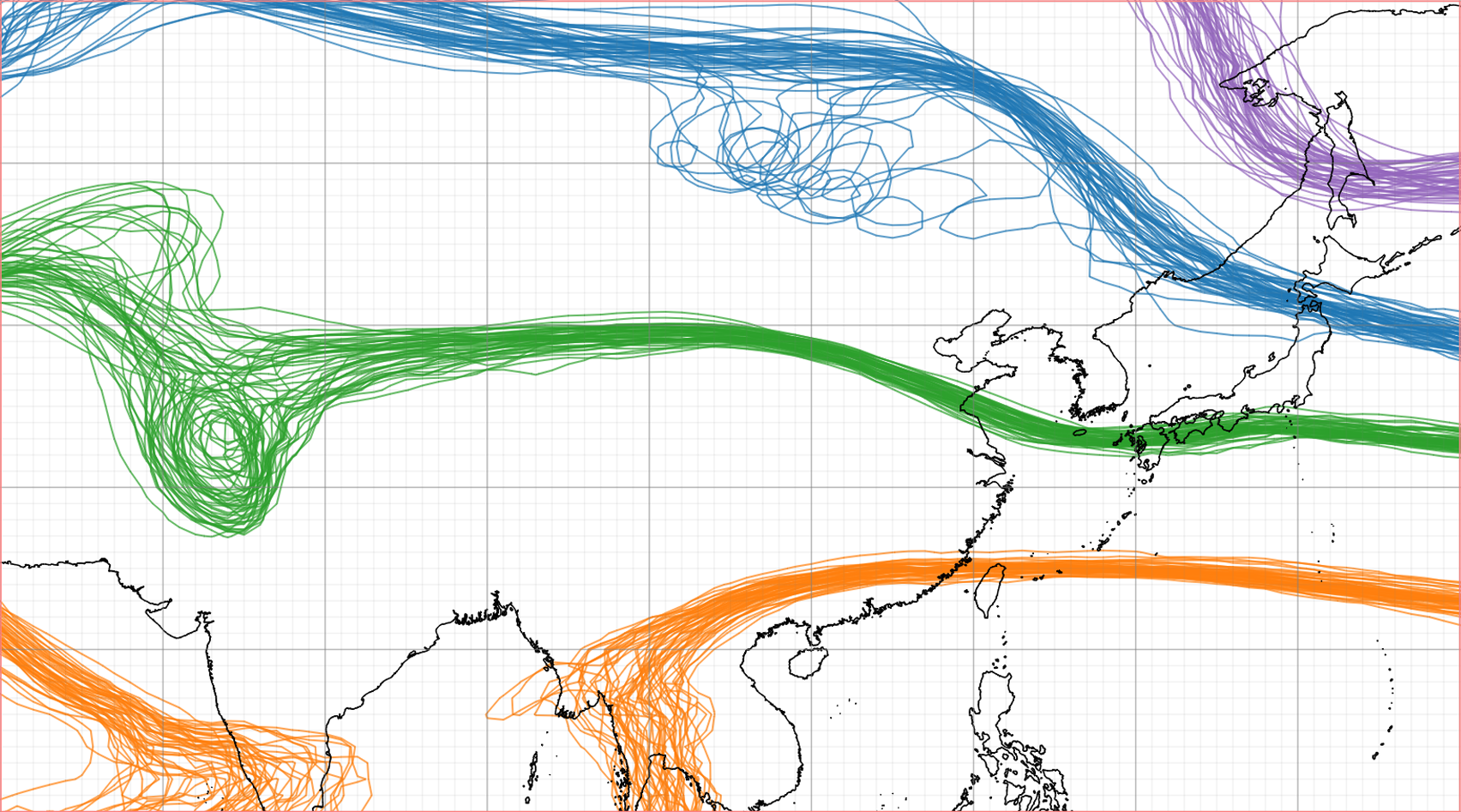}
			\caption{Isolines of global UPs}
			\label{fig:spaghetti_local_g}
		\end{subfigure}
		
		\begin{subfigure}[tb]{0.455\linewidth}
			\includegraphics[width=\linewidth]{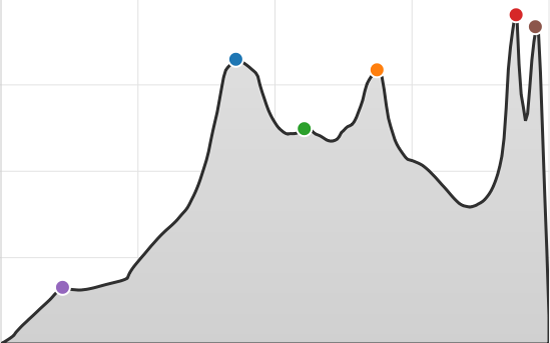}
			\caption{Spreading curve calculated locally}
			\label{fig:curve_local}
		\end{subfigure}
		~
		\begin{subfigure}[tb]{0.51\linewidth}
			\includegraphics[width=\linewidth]{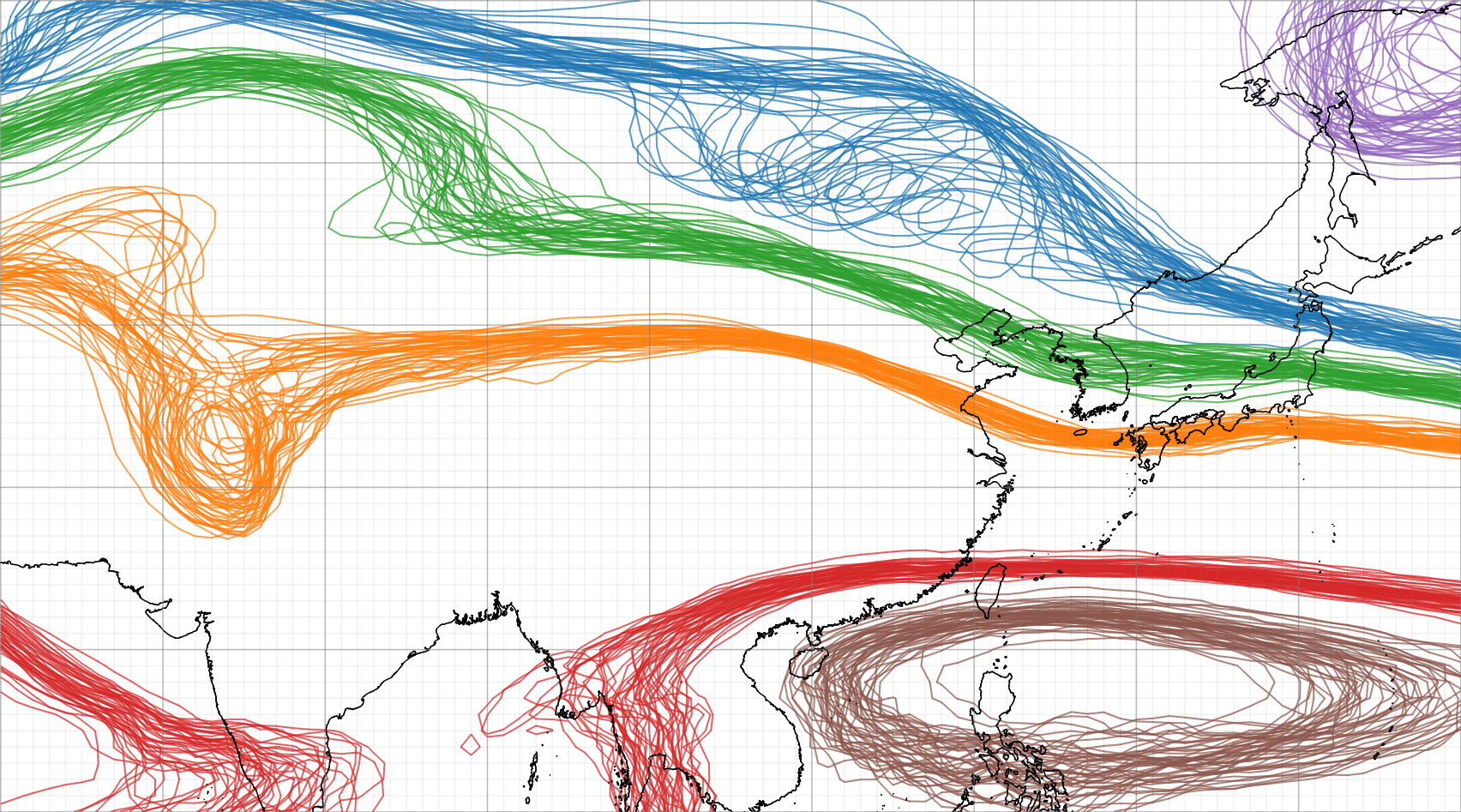}
			\caption{Isolines of local UPs}
			\label{fig:spaghetti_local_l}
		\end{subfigure}
		
		\vspace{-2mm}
		\caption{Variable spreading curves are calculated from the (b) global and (d) local regions. Spaghetti plots of the UPs from the spreading curves are shown for the (a) global and (c, e) local regions, where (c) uses the UPs of the global spreading curve, and (e) uses the UPs of the local spreading curve.}
		\label{fig:global}
		\vspace{-2mm}
	\end{figure}	
}

\newcommand{\regionHeatmapFig}{	
	\begin{figure}[h]
		\vspace{-2mm}
		\centering
		
		\begin{subfigure}[tb]{0.31\linewidth}
			\includegraphics[width=\linewidth]{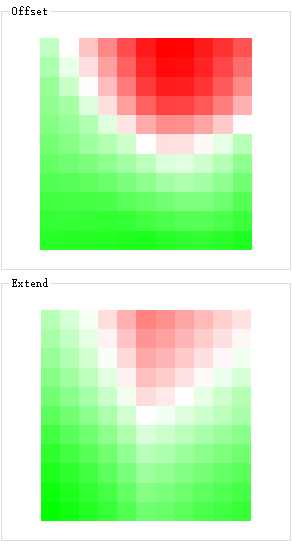}
			\caption{Region A}
			\label{fig:region_heatmap_A}
		\end{subfigure}	
		~
		\begin{subfigure}[tb]{0.31\linewidth}
			\includegraphics[width=\linewidth]{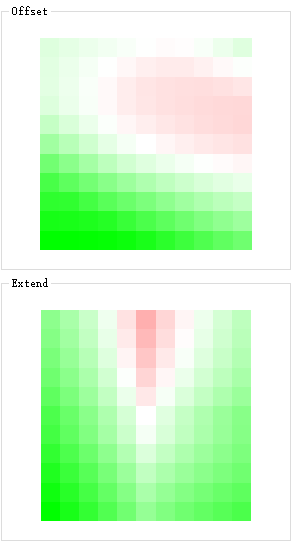}
			\caption{Region B}
			\label{fig:region_heatmap_B}
		\end{subfigure}
		~
		\begin{subfigure}[tb]{0.31\linewidth}
			\includegraphics[width=\linewidth]{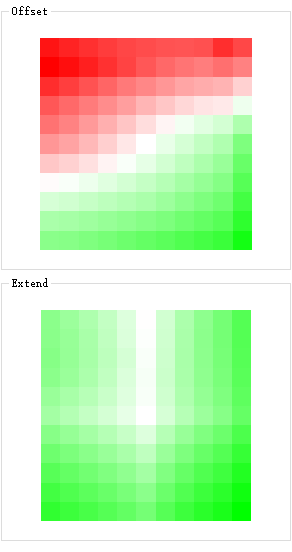}
			\caption{Region C}
			\label{fig:region_heatmap_C}
		\end{subfigure}
		
		\vspace{-2mm}
		\caption{Region stability heat maps of the three regions in Fig.~\ref{fig:meanspread}}
		\label{fig:region_heatmap}
		\vspace{-2mm}
	\end{figure}	
}

\newcommand{\regionAnalysisFig}{	
	\begin{figure}[h]
		\centering		
		\vspace{-2mm}
		\includegraphics[width=\linewidth]{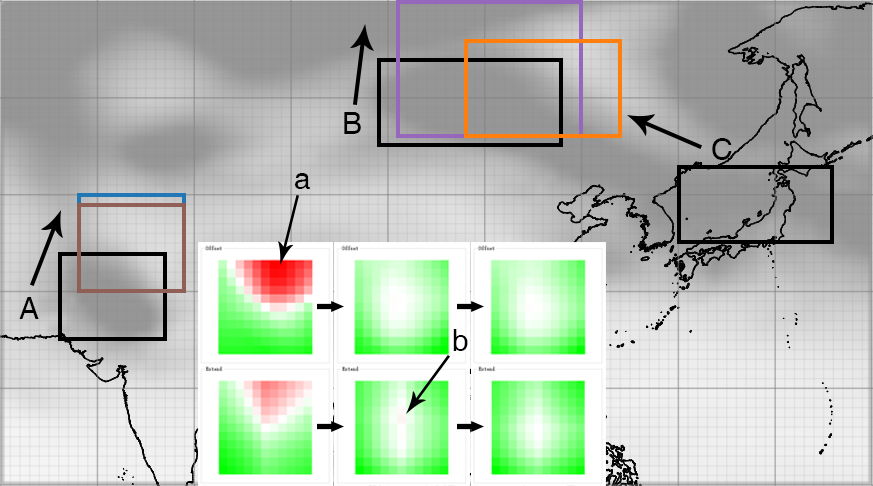}
		\vspace{-4mm}
		\caption{Selected  uncertain regions in Fig.~\ref{fig:meanspread} are adjusted using the assistant heat map. Region A is taken as an example. Two separate clicks on (a) the offset heat map and (b) the extending heat map obtain an optimized result.}
		\label{fig:region_analysis}
		\vspace{-2mm}
	\end{figure}	
}

\newcommand{\temporalFig}{	
	\begin{figure*}[tb]
		\centering		
		\vspace{-2mm}
		\includegraphics[width=\linewidth]{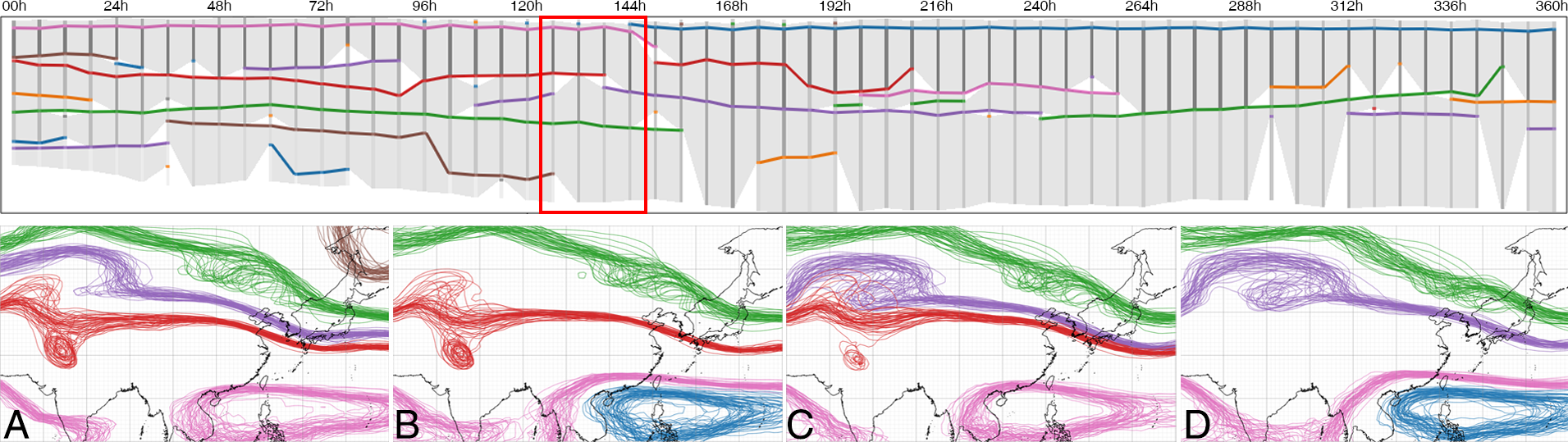}
		\vspace{-4mm}
		\caption{Temporal view and spaghetti plots of four time steps highlighted by a red rectangle. The temporal view shows a red extreme point that ended at 138 h, and the spaghetti plots show the process in which the uncertainty of the red extreme point diminishes.}
		\label{fig:temporal_view}
		\vspace{-4mm}
	\end{figure*}	
}

\frameTeaser

\begin{document}
	\firstsection{Introduction}
	\label{sec:introduction}
	\maketitle
	
	\par As an important method of handling potential uncertainties in modern simulations, ensemble simulation~\cite{bauer2015the,leutbecher2008ensemble} has been widely applied in many disciplines, particularly with the dramatic improvement of modern computing power. Ensemble simulations are conducted multiple times using different models, initial values, and parameters that cover as many distribution spaces as possible; in this manner, incomplete information and insufficient computational accuracy are avoided. However, one resultant issue is that the large-scale ensemble simulation data frequently obscure people's analysis and understanding. Therefore, identifying an appropriate ensemble simulation representation is necessary before conducting further ensemble data analysis.
	
	\par Visualization is a promising and powerful way of representing and analyzing ensemble simulation data~\cite{wang2019visualization}. Although conventional visualization methods demonstrate promising performance in real-world ensemble data analysis~\cite{obermaier2014future,wang2019visualization}, several challenges still exist. \textbf{C.1 Lack of balance between location-based and feature-based analyses.} Obermaier and Joy~\cite{obermaier2014future} classified existing work into two categories, namely, location-based and feature-based methods. Although the existing analyses support selection and exploration on the location and feature dimensions, they are often conducted separately~\cite{obermaier2014future}. However, the two dimensions do not exist independently, i.e., features are distributed in locations, and different locations exhibit differences in features. As indicated by Dasgupta et al.~\cite{poco2014similarityexplorer}, domain experts usually emphasize a certain dimension for spatiotemporal and multimodal simulation result analysis; by contrast, we frequently need to represent both dimensions for improved analysis. In other words, the individual selection and separate analysis of any one of the two dimensions would fail to capture critical elements in the ensemble data and inevitably cause a considerable  loss of information. \textbf{C.2 Insufficient analysis on the temporal dimension.} Previous work either focused on the analysis at one single time step and ignored time-dependent information~\cite{whitaker2013contour}; or oversimplified the details at each time step to cover the analysis at multiple time steps~\cite{fofonov2019projected}. Some work simply combined the results of several time steps together without simplifying which would inevitably introduce visual clutter~\cite{ferstl2017time}. How to represent time-varying ensemble simulation data effectively and sufficiently and uncover underlying patterns and most dominating events~\cite{li2020seevis} remains a great concern for ensemble simulation data analysis. \textbf{C.3 Repeated trial-and-error process.} Existing research provides users with general summaries and representations for subregions~\cite{kumpf2018visualizing} and isovalue selections ; moreover, it displays the details~\cite{ma2019interactive} regarding the spatial and variable dimensions~\cite{potter2009ensemble,sanyal2010noodles}, respectively. Thus, a repeated and inefficient trial-and-error process is required and may easily lead to a significant loss of information. \textbf{C.4 Inefficient intervention.} With the high complexity and multivariance of ensemble data, most existing visualization methods focus on simplifying data to reveal the main structures, e.g., leveraging data depths or probabilistic models to convert contours or curves into confidence bands that can represent uncertainties~\cite{ferstl2016streamline,ferstl2016visual,ferstl2017time,mirzargar2014curve,whitaker2013contour} or adding glyphs to represent uncertainties~\cite{sanyal2010noodles}. However, different sampling or simplification approaches tend to emphasize certain parts of information while ignoring others. In addition, they output a deterministic result, which may lead to cognitive bias without a flexible interaction and intervention mechanism in real-world scenarios.
	
	\mappingFig	
	
	\par To resolve these issues and in consideration of the influence of spatial location, we propose a novel feature-based ensemble data analysis method to achieve a balance between location- and feature-based analyses (\textbf{C.1}). In contrast to previous studies that leverage spatial locations as the primary analysis dimension and perform a global analysis based on the local analysis of each grid point, our approach considers the attribute variables as the primary analysis dimension. We use a simple example for demonstration (Fig.~\ref{fig:mapping}). The data are a two-dimensional scalar field of a single variable, and the data space consists of two dimensions, i.e., location and variable (location is also two-dimensional). Conventionally, the uncertainty is calculated at each grid point according to the ensemble members and then analyzed and visualized separately~\cite{bensema2016modality,gosink2013characterizing}. In our work, we first divide the variable space and then describe its spatial distribution in each corresponding interval to characterize the corresponding uncertainty. This process helps construct an overview of the global uncertainty distribution and assists in the subsequent interactive exploration in the feature space. The comparative procedures that use location and variable as the primary dimension are demonstrated in Figs.~\ref{fig:mapping_location} and \ref{fig:mapping_variable}, respectively. Specifically, the range of the variable is first segmented into bins, and then the spatial distribution of each bin is calculated to generate a variable spreading curve. The spreading curve supports not only the comparison on global and local levels but also the ensemble field and ensemble average field to describe the uncertainty distribution and support abundant interactions for user intervention (\textbf{C.4}). On the basis of this calculation method, we further design and develop a visual analysis framework to help users analyze the ensemble simulation data. Particularly, we propose (1) a spatial distribution curve to facilitate an intuitive understanding of the variable distributions, (2) a stability heat map to understand the uncertainty distribution of the selected region and further guide the modification for an efficient analysis (\textbf{C.3}), and (3) a new temporal view design to help users understand the temporal distribution of features (\textbf{C.2}). The primary contributions of our work are summarized as follows:
	\begin{compactitem}
		\item We propose a novel feature representation approach through calculating the spatial spreading of variables to measure uncertainty distributions, thus balancing analysis on the spatial location and feature dimensions and offering a new perspective for analysis.
		\item We propose an efficient simplification method that represents time-varying ensemble data and conveys underlying patterns.
		\item We design a visual analysis framework to help make choices between spatial and feature dimensions and reduce cognitive biases that may result from separate analysis on a single dimension.
	\end{compactitem}

	\section{Related Work}
	\par Literature that overlaps with this work can be classified into three categories: ensemble visualization, subspace selection in ensemble visualization, and uncertainty measurement and visualization.
	
	\subsection{Ensemble Visualization}
	
	\par Ensemble visualization has been intensively studied in recent years; it has been applied to various domains, such as meteorology~\cite{rautenhaus2018visualization} and ocean-atmosphere research~\cite{afzal2019state}, and various analysis tasks, such as uncertainty~\cite{ferstl2016visual,whitaker2013contour} and parameter analyses~\cite{he2020insitunet,wang2017multi}. In contrast to multivariate and spatiotemporal data, ensemble data introduce a new dimension, i.e., multivalue~\cite{love2005visualizing}, which makes analysis more difficult. Obermaier and Joy~\cite{obermaier2014future} divided ensemble visualizations into feature-~\cite{ferstl2016visual,pfaffelmoser2013visualizing_eurovis,whitaker2013contour} and location-based~\cite{bensema2016modality,gosink2013characterizing,mirzargar2018representative,shu2016ensemblegraph} categories. However, real-world analysis tasks often require a combination of the advantages of the two categories. Wang et al.~\cite{wang2019visualization} surveyed the recent progress on ensemble data visualization and visual analysis; and proposed the challenges of further exploration in variable and ensemble dimensions. Quinan et al.~\cite{quinan2016visually} maintained that the visual inconsistency of different ensemble visualizations makes it difficult for users to establish a mental map. Therefore, they presented several comprehensive improvements. However, the studies above mainly focused on visual designs. Moreover, each method has an interactive mode that is designed for a specific problem; this feature is often inconsistent with other methods, thus causing inconvenience to users. In this study, we provide a new perspective of analysis and particularly focus on the analysis of dimensions of variables, locations, and ensembles. Our proposed framework integrates well-established visualization methods~\cite{ferstl2016visual,pfaffelmoser2013visualizing,pfaffelmoser2011visualizing,pfaffelmoser2012visualization,pothkow2011positional,pothkow2013nonparametric,pothkow2011probabilistic,whitaker2013contour}, thus enabling users to adopt any one of them for comparative analysis. We also provide flexible interactions to help users achieve their analysis goals while maintaining the mental map.
	
	\meanspreadFig	
	
	\subsection{Subspace Selection in Ensemble Visualization}
	
	\par Effective data selection is the basis of further analysis and calculation~\cite{liao2016visualization}. Recently, researchers have developed various interactive mechanisms to support region~\cite{gong2016visualization} and isovalue~\cite{hazarika2018information} selection in ensemble data visualization. For the subregion selection problem, most existing studies provide a rectangular subregion interaction for reanalysis~\cite{potter2009ensemble,sanyal2010noodles}, whereas Holt et al.~\cite{Hollt2013Visual, Hollt2014Ovis} provided an irregular region selection. However, when users conduct exploratory visualization with unclear analysis objectives, irregular region selection could increase interaction complexity. Therefore, in our work, we choose the commonly-used rectangular selection.
	To the best of our knowledge, only a few studies focused on the selection of the attribute variable dimension. Ma and Entezari~\cite{ma2019interactive} proposed a framework to explore ensemble data through interactions. In this framework, users can select and analyze interesting isovalues by displaying the summary contours generated by each isovalue. However, it is still based on some predefined isovalues that largely depend on users' domain knowledge. For the direct selection of isovalues in the variable dimension, methods based on topological characteristics for non-ensemble data have been proposed~\cite{bajaj2009topology}; however, only a few studies focused on ensemble data. In contrast to that of previous research, the method  proposed in the present study assists users in selecting isovalues on the basis of the calculated variable spreading curve. 
	
	\par 
	To find a proper isovalue for contour analysis, Hazarika et al.~\cite{hazarika2018information} calculated two specific information, i.e., surprise and predictability, to assist users in choosing the optimal isovalues. They first assumed the mean field as the ground truth; then, they calculated surprise and predictability between each member and the mean as the description of the ensemble data. Similarly, we calculate value distribution but instead of using specific information for measurement, we find the extreme points of spatial distribution, which do not assume the mean as ground truth making the method more universal.
	
	\subsection{Uncertainty Measurement and Visualization}
	
	\par Uncertainty has been intensively studied in recent years~\cite{liu2019visualizing,yan2020structural,liao2018cluster}. For example, Noodle~\cite{sanyal2010noodles} leverages basic statistics at each grid point as uncertainty metrics. Uncertainties do not always follow the normal distribution; thus, bootstrapping~\cite{efron1994introduction} is introduced for normalization; furthermore, the Kullback-Leibler divergence~\cite{chen2015uncertainty} is introduced to calculate the distance as a supplement to Euler distance because ensemble members may introduce a probabilistic distribution of the data. In our work, we focus on the uncertainty in ensemble data. Therefore, we adopt the idea of sampling and introduce a feature uncertainty measurement method. The calculated similarity is conventionally visualized by line charts to show the trends. For example, Fang et al.~\cite{fang2007visualization} proposed a time--activity curve to show the variance of each voxel and their similarities. Wang et al.~\cite{wang2008importance} divided data into blocks and calculated the importance of each block by using conditional entropy, thus showing the importance over time. These works all leveraged time as the x-axis of the line charts. In our work, we use a line chart to show the calculated feature distribution curve and take the variable as the x-axis to focus on the variance of the features.

\standDeviationFig

	\section{Requirement Analysis}
	\label{sec:requirement}
	
	\par To understand how ensemble data are analyzed in practice, we worked with two senior experts (E.1 and E.2), who have maintained a long-term cooperative relationship with us. We obtained their permission and authorization and conducted a field study to observe their daily practice in meteorological weather forecasting and climate research. A considerable part of their work involves analyzing ensemble data and developing insights accordingly. One major factor that affects the analysis of ensemble simulation results is uncertainty. We interviewed the experts in two separate sessions to identify their primary concerns about ensemble data analysis and potential obstacles to efficiency. The major requirements are summarized as follows:
	
	\par \textbf{R.1 Understand uncertainty distribution from a global perspective.} Conventional visualization methods often leverage color mapping or isocontours to study the uncertainty distribution in an entire scalar field, which frequently introduces misunderstanding. For example, E.1 stated that the effect of uncertainty and the changing gradient of the numerical values are usually intertwined. For example, in Fig.~\ref{fig:meanspread_mean}, the grayscale part of the background is encoded by the ensemble variance, and the contours of the ensemble mean and spaghetti plots are shown in Figs.~\ref{fig:meanspread_mean} and \ref{fig:meanspread_s}, respectively. In the mid-latitude regions, the variance on the east side is considerably higher than that on the west side. However, the expert observed that the isolines in region C are denser than those in the areas with the same latitude in Figs.~\ref{fig:meanspread_mean} and \ref{fig:meanspread_s}. Thus, he concluded that the high variance in region C is caused by the high gradient rather than the uncertainty. In region A, the spaghetti plots show that the contours of different isovalues pass through this region, indicating that the high variance is caused by uncertainty. However, upon further investigation, the expert found that the most crossed point does not coincide with the highest variance, whereas the latter is coincident with the compact contours, indicating that the high variance in region A is caused by uncertainty and high gradient. In region B, the high variance fits perfectly with the uncertainty shown by the spaghetti plots. This observation indicates that the variance could be caused by different factors. Therefore, E.1 required to show the distribution of the ensemble data and the uncertainty simultaneously so that they would not interfere with each other.
	
	\par \textbf{R.2 Discover key points in variable space.} The use of isocontours to visualize the scalar field requires appropriate isovalues to be predetermined; this requirement is difficult to satisfy. As shown in Fig.~\ref{fig:meanspread_s}, visualizing $11$ isovalues by using a spaghetti plot causes visual clutter and misses important features. However, if the preset isovalues increase with $30$ (kpm), then an interesting structure of the pink isolines appears (Fig.~\ref{fig:meanspread_s_2}). However, this operation largely depends on domain expertise. Thus, the experts required having a mechanism that can automatically discover and recommend interesting structures. 
	
	\par \textbf{R.3 Identify region locations and ranges with uncertainty.} For a detailed analysis, the experts first locate the areas with large uncertainty and then determine the corresponding ranges. However, identifying the regions is relatively easy, and determining the corresponding ranges difficult due to the complexity of ensemble simulation. The experts must select different regions manually and repeatedly; in this manner, the analysis target is identified albeit time-consuming.
	
	\par \textbf{R.4 Support temporal analysis of the ensemble data.} In some scenarios, such as weather forecasting, temporal analysis can uncover the underlying patterns and interesting regions on the temporal dimension. However, without an efficient simplification of handling the time-varying ensemble simulation data, conducting temporal analysis thoroughly would not be feasible. Data analysis with one single time step can be complicated enough, let alone multiple time steps. Therefore, finding an efficient method to simplify the time-varying ensemble data and extract important features is a great concern for the experts.

	\par \textbf{R.5 Provide interaction mechanisms in feature space.} 
	The existing ensemble visualizations for the representation of ensemble data features are often based on preset conditions or parameters. For example, the isoline display frequently relies on preset isovalues. However, according to our experts, these preset conditions do not always meet their requirements in real-world scenarios because ``\textit{the analysis goals differ in various scenarios.}''  Therefore, our experts require improved interaction mechanisms to support their flexible selection in the feature space.

	\section{Uncertainty Distribution via Variable Spreading}
	
	\par To meet the above requirements for describing the distribution on the location and feature dimensions, we must define a method to measure the uncertainty distribution. In the following subsections, we introduce the relevant terms, describe our proposed methods for the uncertainty calculation, and introduce our visual analysis framework.

	\subsection{Definition of Variable Spreading}
	\label{sec:spread_def}
	We first introduce the definition of variable spreading. 
	As mentioned in Section~\ref{sec:requirement}, the high variance demonstrated by the density map may be caused by a high gradient rather than the uncertainty. In addition, the experts stated that the effect of the uncertainty and the changing gradient of the numerical values are usually intertwined due to the correlation between the variable dimension and the ensemble member distribution. Here, we define \textbf{gradient} as the change rate of attributes in the spatial dimension and \textbf{spreading} as the spatial region that the attributes value cover. Suppose a scalar field defined by function \(f: p(x,y) \to v\), \(p(x,y)\) is a point in the spatial region and \(v\) is the scalar value of the variable at this point; therefore, the gradient at this point is defined as \( \nabla f(p)=\{\frac{\partial f}{\partial x},\frac{\partial f}{\partial y}\}\).
	\par We showcase the concept of variable spreading through a simple example, which contains only two ensemble members that are functions defined on the 1D space (Fig.~\ref{fig:spreading12}). One of the members (i.e., green curve) is defined by the function \(v=x^2\), and the red one is defined by the function \(v=(x-0.1)^2\). According to the definition, for any two points on the Y axis, the corresponding deviation of the $x$ coordinate of the two members are constant, i.e., the two red segments $a$ and $c$. However, for any different points on the X axis, their corresponding deviation of the $y$ coordinates varies (i.e., the two pink segments $b$ and $d$). This observation can be also reflected by the curve of the standard deviation. In this example, the variable spreading of v1 is $c$ , and the variable spreading of v2 is $a$.  However, in this case, we can maintain that the uncertainty is uniformly distributed and thus inconsistent with the color mapping of the standard deviation. Nevertheless, it offers practical guidance. For example, in the field of meteorology, the prediction of a certain weather system, e.g., a high-pressure air mass, may have spatial deviation due to the different members of the ensemble.
	
	\par We also use 1D data to verify the efficacy of our proposed measure. We have already shown the case in which uncertainty is equally distributed but the standard deviation is not consistent in the last paragraph. Now we add a new member defined by the following function:
	\begin{displaymath}
	v=\Big\{
	\begin{array}{cc}
	(x-0.5) \times 0.8+0.6 & x>0.5 \\
	(x-0.5)/0.8+0.6 & x\le 0.5 \\
	\end{array}
	,
	\end{displaymath}
	which is the first member transferred in the X-axis with scaled distance getting the maximum value when \(x=0.5\). Thus, when the ensemble contains $m1$ and $m3$, the maximum of uncertainty locates at the point of \((0.5,0.36)\). Both the spreading and the standard deviation can reflect this phenomenon (Fig.~\ref{fig:spreading13}). However, the gradient change brought about by adding $m2$ masks the uncertainty that can be previously reflected by the standard deviation (Fig.~\ref{fig:spreading123}).
	
	\par Our proposed variable spreading can be regarded as the $x$ range covered by different ensemble members of one point in the Y axis, which is the axis of the variable value. In the two-dimensional space, it corresponds to the area of the spatial region where certain values fall in the range of the values of the ensemble members. If the functions of \(m\) ensemble members are \(f^{(i)}: D \to R, i\in[1,m]\), then the spreading is defined by:
	\begin{equation}
	\pi_f(v)=\int_D{\delta \left(\min_{\forall i \in [1,L]}\{f^{(i)}(x)\}\le v \le \max_{\forall i \in [1,L]}\{f^{(i)}(x)\}\right)dx}
	,
	\end{equation}
	where \(\delta\) is the continuous Dirac delta, which is defined as:
	\begin{equation}
	\delta(x)=\Big\{
	\begin{array}{cc}
	\infty & x=0 \\
	0 & otherwise \\
	\end{array}
	.
	\end{equation}

	\subsection{Variable Spreading Calculation}
	\label{sec:variable_spreading_calculation}
	\par We propose a novel variable spreading procedure and calculate it through the following three steps. Suppose the ensemble contains \(L\) members. First, the range of the values of the entire ensemble data \([V_{min},V_{max}]\) is calculated. Second, the range is divided into \(N_{iso}-1\) bins to generate \(N_{iso}\) isovalues \(I_1,...,I_{N_{iso}}\), where \(N_{iso}\) is preset by the user in accordance with the specific analysis requirement. The finer the analysis, the more bin it takes. Third, we calculate the spreading of each isovalue \(I_i\) in the spatial region, denoted as \(H(I_i) \in [0,1]\), which is then used to describe the feature distribution in space.
	
	\par 
	Our calculation method for the spreading of each isovalue \(I_i\) is based on sampling points. Specifically, \(N_s\) points \(P_1,...,P_{N_s}\) are sampled in the spatial region. For each point \(P_j, j \in[1,N_s]\), the corresponding attribute values denoted as \(v_j^{(1)},...,v_j^{(L)}\), in all the ensemble members are calculated via bilinear interpolation. The range of ensemble values at this point can then be calculated as \(C_j=[\min_{\forall k \in [1,L]}v_j^{(k)}, \max_{\forall k \in [1,L]}v_j^{(k)}]\), and the spreading of each isovalue \(I_i\) can be defined as:
	\begin{equation}
	\label{eqn:h}
	H(I_i)=\frac{\sum_{j=1}^{N_s}{\delta_{I_i}(C_j)w_j}}{\sum_{j=1}^{N_s}{w_j}},
	\end{equation}
	where \(\delta_{I_i}\) is a Dirac measure defined as:
	\begin{equation}
	\delta_x(A)=\Big\{
	\begin{array}{cc}
	1 & x \in A \\
	0 & x \notin A \\
	\end{array},
	\end{equation}
	\(w_j\) is the weight of the sampling point, which is calculated by the map projection.
	\(w_j=\cos \theta_{lat}\), where \(\theta_{lat}\) is the current latitude of the sampling point. 
	The geodesic distance of different latitudes varies; thus, we use this weight to eliminate this deviation. The obtained \(H(I_i)\) represents the percentage of the relevant regions of \(I_i\) in the spatial region. When all the results of the \(N_{iso}\) isovalues are obtained, they constitute the spreading curve of the variable.

	\subsection{Variable Spreading Curve}
	\par After calculating the spreading of each isovalue, a spreading curve is established along the range of the variable. These curves can represent the uncertainty distribution of the variable. However, three problems remain if we leverage these curves to represent the uncertainty directly.
	
	\par \textbf{P.1 Effect of the spatial distribution of the data itself.} 
	The variable is not evenly distributed in the scalar field. Thus, the calculated spatial spreading curve is affected not only by the uncertainty but also by the spatial distribution of the data. For example, values that are in the border of the region cover few spatial areas even with large uncertainty. Therefore, the effect of the uncertainty and of the original data distribution should be decoupled first before further analysis.
	
	\par \textbf{P.2 Local region uncertainty covered by the global region uncertainty.} 
	As the spatial spreading curve is calculated globally, the uncertainty of the local region may be hidden by the global uncertainty distribution, thus requiring an analysis interaction to support different levels of detail (LOD).
	
	\par \textbf{P.3 Region selection.} An inappropriate selection of a region generates a biased and/or misleading result. Moreover, region selection requires domain knowledge. Visual cues are still required because even professional meteorological researchers cannot easily identify the most appropriate analysis areas when dealing with a new ensemble simulation dataset.
	
	\par To resolve these problems, we design an interactive visualization system based on the spreading curve coupled with different visualizations and interaction mechanisms to enable the analyzers to explore the ensemble simulation data further. We first summarize how we deal with the above three problems.
	
	\subsubsection{Decoupling Between Uncertainty and Data Distribution}
	\par To decouple uncertainty from data distribution, data distribution should be measured first. Average scalar field of the ensemble can be used. Researchers have proposed many complicated measures, such as Bayesian model averaging~\cite{gosink2013characterizing} and voting~\cite{liao2015visual}. Here, we use the ensemble mean at each grid to establish the average field. As shown in Fig.~\ref{fig:parameterA_0}, the filled region is the calculated spatial spreading curve, whereas the blue curve is the spreading curve in the average field. A correlation can be observed between the two curves. For each bin in the range of the variable, the calculated spreading value in the ensemble field and in the average field is \(s_i\) and \(\bar{s_i}\), respectively. A straightforward method to use their quotient to represent uncertainty. Fig.~\ref{fig:parameterA_1} shows the curve of the quotient. Overall, the two ends are highlighted while the middle area presents a single peak distribution.
	
	\parameterAFig
	
	\par Considering that completely removing the effect of the spatial distribution of the variables is unreasonable and that variables with a small spatial distribution but large uncertainty have little analysis value, we introduce a parameter \(\alpha \in[0,1]\) to coordinate the relationship between the two and obtain the final variable spatial distribution measurement as follows:
	\begin{equation}
	\label{eqn:alpha}
	\tilde{s_i}=\frac{s_i}{\bar{s_i}^\alpha}.
	\end{equation}
	\par We can then easily determine that \(\tilde{s_i}=s_i\) when \(\alpha=0\), thereby indicating that the effect of the data distribution is totally ignored; in addition, \(\tilde{s_i}=\frac{s_i}{\bar{s_i}}\) when \(\alpha=1\), thereby indicating that the effect of data distribution is fully considered.

	\subsubsection{Region Analysis}
	\par To solve the problem of local uncertainty being covered by the global calculation (\textbf{R.3}), we provide an LOD interactive exploration of the selected region. To avoid a blind choice of the regions, we offer three mechanisms - region recommendation, region stability analysis, and region selection guidance - to help analysts improve their selections. Region recommendation involves recommending the initial analysis region in accordance with the preset rules, region stability analysis is performed to evaluate and display the region's stability after the analyst selects a region, and region selection guidance provides analysts with modification guidance on the basis of the region's stability display.
	
	\subsubsection{Region Stability Visualization}
	\label{sec:region_stability}
	
	\par Latter analysis is severely affected by the region selection, which depends on the domain knowledge and experience of the analyst. However, even the most experienced analyst cannot guarantee the right choice of region. Furthermore, region choices differ in various circumstances and analytical tasks. Therefore, showing the stability of the region is critical (\textbf{R.3}). The objective of our analysis is uncertainty; thus, we define the stability of a region as the stability of its uncertainty, which is measured by comparing with the regions after translation and expansion.

	\regionComparisonFig
	
	\par To compare spreading curves, we overlay and display the spatial spreading curves of the regions which are the selected region biased in four directions to show the stability of the selected region. As shown in the first column of Fig.~\ref{fig:region_comparison}, the green, red, orange, and purple curves are the spatial spreading curves in the regions corresponding to the selected region biased in the left, up, right, and down directions by one grid, respectively. In Fig.~\ref{fig:region_comparison_spaghetti_1}, the bottom of an uncertain region is selected. In its corresponding biased curves view (Fig.~\ref{fig:region_comparison_curve_1}), the red curve has a higher peak, while the purple curve has a lower one, indicating that more complete uncertainty regions would be covered by moving the region upward. The bottom-right corner of the uncertainty region is selected in Fig.~\ref{fig:region_comparison_spaghetti_2}, and the curves moving left (green) and upward (red) have higher peaks. There are also cases with multiple peaks (Fig.~\ref{fig:region_comparison_curve_3}), showing that more than one direction bring uncertainty, i.e., the region is in the gap between different uncertainty regions.

	\par Considering using only one mesh offset could be accidental, we calculate the spreading curves for four directions with 1-5 offsets. However, $100$ spreading curves are generated which would cause severe visual clutter if displayed simultaneously. After a careful observation and analysis of these curves, we decide to use the maximum to represent the curve. Therefore, the change of uncertainty for each biased region \(D_{mn}\) is calculated by formula:
	\begin{equation}
	Q(D_{mn})=\max_{\forall I_i} H_{D_{mn}}(I_i)-\max_{\forall I_i} H_{D}(I_i),
	\end{equation}
	where \(H_{D_{mn}}(I_i)\) is calculated using formula~\ref{eqn:h} in region \(D_{mn}\), which is region \(D\) biased by \(m,n \in [-1,5]\) grid point in the horizontal and vertical direction. In order to display in the same scale, we further calculate the uncertainty change ratio through normalization by the biggest absolute change value:
	\begin{equation}
	R(D_{mn})=\frac{Q(D_{mn})}{\max_{\forall l,s\in[-5,5]}| Q(D_{ls})|}.
	\end{equation}
	As \(R(D_{mn})\in [-1,1]\),we can use color mapping to show it. As shown in the third column of Fig.~\ref{fig:region_comparison}, red means positive, and green means negative, and the saturation of the color maps to the absolute scales from \((0-1)\). The figure shows that the heat map is continuous and consistent with the previous conclusion, thereby providing more comparative migration results while the biased curve shows more detailed migration details. The two views jointly provide the analyst with region stability and guidance for region selection.

	\subsection{Extreme Points and Feature Space Segmentation}

	\par After calculating the spreading of each isovalue, we visualize the spreading curve in Fig.~\ref{fig:frame}(E), from which we can observe that the curve is not smooth but has several peaks and valleys. These extreme points indicated some potentially important characteristics.
	
	\par The two kinds of extreme points (i.e., maximum and minimum points) have different physical meanings. The maximum points, which have a higher spreading value compared with their neighbors, are considered points with more uncertainty. After discussing with the domain experts, we knew that the large value of uncertainty often indicates the interaction of two meteorological systems. Therefore, we refer to these points as uncertainty points (UPs). The minimum points, which have a lower spreading value compared with the neighbors, are considered points with a small value of uncertainty, often indicating the centers of the meteorological systems; we call these centers stable points (SPs) because of their stability. We then use these points to generate the initial isovalues for users, as shown in Figs.~\ref{fig:stable_points} and \ref{fig:uncertainty_points}.

	\criticalPointFig
	
	\par In Fig.~\ref{fig:critical_points}(a), we show the spaghetti plots of three SPs; the blue one and the green one show a higher degree of unity, whereas the purple one spreads within a small region in the corner. Further inspection of the spreading curve in Fig.~\ref{fig:frame}(B) shows that the SP of the purple spaghetti plot is not remarkable and may be considered a fake SP. In Fig.~\ref{fig:critical_points}(b), we present the spaghetti plots of four UPs, which show significantly higher uncertainty compared with those in Fig.~\ref{fig:critical_points}(a), i.e., one meaningless curve in the corner and three curves worth further analysis. The distinction between UPs and SPs can help users find the uncertainty distribution in the variable and spatial dimensions (\textbf{R.2}).

	\subsection{Parameter Discussion}
	\par During the calculation of the isovalue spreading curve and the extreme points, different parameter settings exhibit various effects on the corresponding results. In this section, we discuss different parameter choices and their corresponding effects on the results. 
	
	\par As described in Section~\ref{sec:variable_spreading_calculation}, we calculate the spreading by using sampling points. Two alternative approaches, namely, random and regular sampling, are used to obtain the sampling points. Theoretically, the results of both methods tend to represent the results accurately when the number of sampling points tends to infinity. We try sampling points of \(1,2,4,8,16,32,64\), and \(128\) times of the grid points, use both the results of 128-times grid points with these two methods as the baseline, and calculate the average bias of the curve. The results are shown in Fig.~\ref{fig:sampling_points}. From the bar chart, we can observe that the regular sampling point method demonstrates an improved result, except for the first column, which leverages the number of grid points as the number of sampling points and whose result is too coarse for calculation. Therefore, we choose regular sampling. Fig.~\ref{fig:sampling_points} also shows that 16-times grid points are the elbow point; thus, we choose this number of sampling points to balance the calculation efficiency and effect.

	\samplingPointsFig
	
	\par Considering the computing performance and focusing on the overall trend instead of the details, by default, we set the parameter \(\alpha\) to 0 in Equation~\ref{eqn:alpha} when calculating the variable spreading and overlaps between time steps. However, when analyzing a specific time step, users can manually adjust the parameter. Initially, we determine the extreme point where the spreading value is higher or lower than that of its two adjacent points. However, this approach inevitably leads to too many extreme points. Experimental observation shows that most of the obtained extreme points are due to random jitters of the underlying data. Therefore, we add interval smoothing, which considers a point as extreme only when it is the maximum or the minimum value among its neighbors with radius \(\beta\). In our case, we set \(\beta=5\) as the default value, and it can be easily adjusted to adapt to different scenarios.
	
	\section{Visualization Design}
	\par Based on the proposed feature spreading technique, we build an interactive visualization framework (Fig.~\ref{fig:frame}), which consists of (A) a parameter setting panel, (B) a region stability heat map, (C) a 2D map view, (D) a temporal analysis view, (E) a spreading curve view, and (F) a display control toolbar. Particularly, the toolbar helps switch among global visualization methods. However, for more complicated operations, e.g., turn off or add a new isovalue, users have to use the statistical view so that they can choose isovalues from the feature distribution curve directly (\textbf{R.3}).
	
	\subsection{2D map view}
	
	\par Following conventional practice, we use a 2D map view to display the overall information (\textbf{R.1}). In addition to providing straightforward ensemble visualization methods, such as spaghetti plots and color mapping in the 2D view, other well-established ensemble visualization methods, such as contour boxplots~\cite{whitaker2013contour}, variability plots~\cite{ferstl2016streamline,ferstl2016visual,ferstl2017time}, and contour probability plots~\cite{kumpf2018visualizing} are available. Furthermore, to support the selection of subregions in the location-feature space (\textbf{R.3}), we leverage the design rationale of ``link + view'' by coordinating other views with the 2D map view.
	
	\subsection{Spreading Curve View}
	
	\par Line charts can intuitively display the one-dimensional distribution of data and compare multiple distributions and thus satisfy our requirement (\textbf{R.2}). Particularly, we use two line charts to show the global and local space spreading. The former focuses on displaying global information and the relationship between uncertainty and data distribution, whereas the latter focuses on local information and region stability.
	
	\par In the global spreading line chart, we show the spatial spreading curve in the average field and the local spatial spreading curve for comparison. The results of adjusting parameters \(\alpha\) and \(\beta\) in the parameter control panel are also reflected in the view. The line chart of the spreading curves of the selected region is mainly used to highlight the spatial spreading of the selected region. The curve of the selected region after moving one grid in four directions is displayed together with the curve of the original region help analysts further understand the stability of the region. Additional offset curves can also be shown by configuring the parameter setting panel. We show only four lines by default to avoid visual clutter. 
	
	\subsection{Region Stability View}
	
	\par As discussed in Section~\ref{sec:region_stability}, 
	we use the offset line chart and stability heat map to show the stability of the selected region (\textbf{R.3}). The offset line chart and the stability heat map can support three kinds of region stability calculation, namely, offset, extend, and shrink. However, the experts maintained that shrink is trivial because the selected region has already been selected and is small enough. Therefore, offset and extend are enough for the stability analysis. The calculation of the stability heat map requires the histogram of the selected regions. To speed up on-the-fly calculation and achieve real-time interactions, we leverage the integral histogram, which was first proposed by Porikli~\cite{porikli2005integral} and has been applied to many follow-up studies~\cite{lee2013efficient,mei2020rsatree}.

	\subsection{Temporal Analysis View}
	\par Our proposed spatial coverage curve offers a new way to handle time-varying ensemble data representation and analysis, as well as temporal changes. Specifically, the variable domain of each time step is segmented in accordance with the extreme points in the spatial coverage curve. The changes of these segments over time reflect the temporal changes of the uncertain spatial distribution.
	
	\par \textbf{Temporal analysis.} The discussion with the experts indicates that if the analysis target is the attribute itself, then the fixed isovalues can further reflect the change of the attribute. However, if the analysis target is the ensemble uncertainty, then the fixed isovalues may fail to reflect those critical aspects due to the effects of the attribute trends over time. Our approach handles this problem by using the dynamic extreme points instead of the fixed isovalues. The continuous forecasting procedure assures continuous extreme points and thus sufficient to bring additional and clear dynamic changes to the analysts.  However, establishing the matching of the extreme points between adjacent time steps is important because the forecast data are not continuous but based on time steps with fixed space. 
	
	\par \textbf{Extreme point alignment.} We build the matching of extreme points between adjacent time steps on the basis of the matching of the segments. For two adjacent time steps \(t_i\) and \(t_{i+1}\), their variable ranges are segmented into a list of segments \(S_{i}^{(1)},...,S_{i}^{(C_i)}\) and \(S_{i+1}^{(1)},...,S_{i+1}^{(C_{i+1})}\) according to their corresponding extreme points \(CP_i^{(1)},...,CP_i^{(C_i-1)}\) and \(CP_{i+1}^{(1)},...,CP_{i+1}^{(C_{i+1}-1)}\). The correlation matrix of the two time steps can be calculated as
	\begin{displaymath}
	\left( \begin{array}{cccc}
	c_{11} & c_{12} & \ldots & c_{1C_i} \\
	c_{21} & c_{22} & \ldots & \ldots \\
	\vdots & \vdots & \ldots \\
	c_{C_{i+1}1} & c_{C_{i+1}2} & \ldots & c_{C_{i+1}C_i} 
	\end{array} \right),
	\end{displaymath}
	where \(c_{ij}\) is calculated by the percentage of the sampling points that fall in both segments. Under the premise that no crossing occurred, we choose the maximum value to establish the correlation between the two time intervals.
	
	\par After the correlation between the segments has been established, we build the connection between the extreme points as follows:
	\begin{compactitem}
		\item If \(S_{i}^{(j)}\) and \(S_{i+1}^{(k)}\) are correlated and \(S_{i}^{(j+1)}\) and \(S_{i+1}^{(k+1)}\) are correlated, then we assume \(CP_{i}^{(j)}\) and \(CP_{i+1}^{(k)}\) are correlated.
		\item If \(S_{i}^{(j)}\) and \(S_{i+1}^{(k)}\) are correlated and \(S_{i}^{(j+1)}\) and \(S_{i+1}^{(k+l)}\) are correlated, then we choose from \([CP_{i+1}^{(k)},CP_{i+1}^{(k+l-1)}]\) the one that is closest to \(CP_{i}^{(j)}\) as the correlated one.
		\item This calculation is performed sequentially from the first time step. For any time step  \(t_i\), if \(CP_i^{(j)}\) is correlated with \(CP_{i-1}^{(k)}\), and the label of \(CP_{i-1}^{(k)}\) is set to \(CP_i^{(j)}\); otherwise, a new label is set to \(CP_i^{(j)}\). To avoid confusion, the newly assigned label does not use the one that appeared in the previous time step.
	\end{compactitem}
	\par After assigning a label to each extreme point, their corresponding visual elements are drawn with the same color to keep the mental map.
	
	\par \textbf{Visual design.} We design a temporal analysis view to show this change, as shown in Fig.~\ref{fig:frame} (D). Each of these columns represents a time step, reflecting the time passing from left to right. In each column, the value range distribution of variables lies from the bottom to the top, and the value range is segmented in accordance with the extreme points. The color of each segment reflects the uncertainty calculated in accordance with the spatial distribution of the segment. The color of the lines between the segments is consistent with the other visual elements of this extreme point, e.g., the spaghetti plots and the extreme point in the spreading curves. Lines with the same color in adjacent columns jointly show the trends. Overall, this view can intuitively show the data change over time and help quickly locate the exceptions.

	\subsection{Interaction among the Views}
	\par In addition to the defining capabilities of our framework, rich interactions are integrated to catalyze an efficient in-depth analysis. \textit{(1) Clicking and brushing.} We provide brushing operations in the 2D view to enable users to observe the distribution of features in a subregion and compare it with the global distribution. In practice, we have also tried other more complicated region selection mechanisms, such as multipoint contour selection or Boolean operation of multiple selected regions. However, the preliminary feedback from the experts indicates that these methods are not applicable. Ultimately, we adopt the simple rectangular selection. In some cases where data distribution cannot be covered by standard rectangles, we rely on real-time user feedback, which determines the areas through multiple choices of observation changes. This interaction helps users easily locate areas with uncertainties (\textbf{R.2}). \textit{(2) Linking among the views.} When a subregion is selected, our framework calculates the feature spreading of the region and compare it with that of the entirety. As shown in Fig.~\ref{fig:frame}, when a subregion with high uncertainty is chosen, e.g., $a$ in $C$, the framework displays the spreading curves of this subregion. This interaction helps establish the relationship between the two dimensions and obtain the global information(\textbf{R.1}). \textit{(3) Selecting points of interest on the curve.} Our framework also supports finding points of interest on the curve and displaying the corresponding spaghetti plots in the 2D map view. In contrast to conventional selection by using predefined isovalues, this interaction expands the analysis scale in the variable dimension (\textbf{R.5}). \textit{(4) Region adjustment by the stability heat map.} Heat maps are purely used for display in previous applications. Here, we use it innovatively for interaction. As each grid in the heat map represents the uncertainty of a biased region, we can switch to that region by directly clicking this grid. This kind of interaction can combine visualization and interaction intuitively, combine with context organically, and provides great convenience for the choice of region (\textbf{R.3}). On the basis of an analysis of the requirements of the analysts, we provide two interaction modes: region selection and region analysis. In the region selection mode, the displayed features do not change with the change of region selection and always retain the global features; thus, analysts can conveniently evaluate the selected regions. When the analyst is satisfied with the selected region, he can choose to switch to the region analysis mode. At this time, the feature is the local feature calculated in accordance with the selected region. The analyst can still adjust the region in the analysis mode. The system assumes that the analyst has established the mental map of global characteristics at this time and then analyzes different regions.

	\globalFig
	
	\section{Evaluation}	
	\par We demonstrate how the domain experts leveraged our system to analyze a real-world ensemble weather forecast dataset adopted from the European Center for Medium-term Weather Forecasts. The datasets are the forecast data from 0:00 on January 1, 2019, which has a total of $50$ members with a forecasting interval of $6$ hours and $60$ time steps.
	
	\subsection{Identifying the Most Remarkable Features}	
	
		\regionHeatmapFig
		
	\par After loading the dataset, the experts wanted to have an overview of the dataset. They have already applied the same data and time steps as shown in Fig.~\ref{fig:meanspread}; hence, they used the isovalues of the UPs calculated by our method. Fig.~\ref{fig:spaghetti_global} displays the spaghetti plot of the global region. The above-mentioned special structure is clearly visualized in the region highlighted by a red rectangle, indicating that the spreading curve can automatically identify the most significant features.

	\par The experts wondered whether the recommended features would change in accordance with different regions and scales; thus, they selected a region of interest. The corresponding spreading curve is shown in comparison with the global curve (Fig.~\ref{fig:curve_global}). However, using the aspect ratio of the global curve does not present an intuitive result; therefore, the experts switched to the local analysis mode. A new spreading curve is shown by considering this region as a global one (Fig.~\ref{fig:curve_local}), and the spaghetti plot is also regenerated by using the UPs of this curve as isovalues (Fig.~\ref{fig:spaghetti_local_l}).

	\par After comparing the two spreading curves, the experts found that the local curve has additional fluctuations and UPs. This finding is consistent with the analysts' experience, i.e., additional features can be identified in the context of small regions; these features are ignored in larger regions.  With a further comparison between the spaghetti plots (Figs.~\ref{fig:spaghetti_local_g} and \ref{fig:spaghetti_local_l}), the experts found that the two plots are similar, except for the extra isovalues in the global spreading curve, indicating that the calculated features are consistent. This observation is somewhat unexpected because the experts had no idea that the global and local calculated features are always like this. Given this unexpected result, they could analyze global consistency. 
	
\regionAnalysisFig

\temporalFig

	\subsection{Region Analysis}
	
	\par After comparing the Asian and global regions, the experts wanted to analyze additional specific regions further. Their first impulse was to evaluate the selected regions, as shown in Fig.~\ref{fig:meanspread}. Particularly, they selected the three regions separately and checked the corresponding region stability heat maps, as shown in Fig~\ref{fig:region_heatmap}.

	\par The offset heat map of region A indicates that the direction which has the most significant increase in uncertainty is east by north; this finding is consistent with the experts' experience mentioned in Section~\ref{sec:requirement}. The extending heat map indicates the same direction but with less strength. This observation shows that compared with the extending map, regional offset could introduce more uncertainty. Regarding region B (Fig.~\ref{fig:region_heatmap_B}), both offset and extending are weaker than those of region A. This finding is also consistent with the experts' experience mentioned in Section~\ref{sec:requirement}, i.e., region B is a relatively certain uncertainty region. Moreover, extending brings a greater increase in uncertainty compared with offset. In region C (Fig.~\ref{fig:region_heatmap_C}), the heat map indicates a different uncertainty direction.

	\par The experts then adjusted the regions by clicking the reddest positions in the heat map to find the regions with the highest uncertainty. The procedure for region A and the results of all three regions are shown in Fig.~\ref{fig:region_analysis}. For region A, the experts clicked on the heat map (position b), which brings the selection to the north-east brown rectangle, where the heat maps show a slight amount of red. They then clicked on the extending heat map (b), which extends the region to the north by two gridlines, thus reaching stability (the extended part is visualized in blue). The purple and orange rectangles indicate the final regions B and C, respectively. The experts witnessed that a large part of the two rectangles overlaps, indicating that they converge to the same region. However, the differences between the two regions show that the final suggestion is affected by the initial selection. In other words, no global optimal choice is available, and it needs to be initialized in accordance with the analysts' requirements.

	\subsection{Temporal Analysis}
	\par After finishing the analysis at one static time step, the experts continued to investigate the continuity of their findings in the temporal dimension. They first checked the temporal view, as shown in Fig.~\ref{fig:temporal_view}, and observed that the division of each time step tends to become increasingly simpler over time in general. This finding is consistent with the fact that prolonged time corresponds to an increased forecast uncertainty and thus fuzzier features. The experts identified a disconnection of the red line at $138$ h; at the same time step, the purple line appears close to the red line. The experts wondered whether this phenomenon is caused by mismatching; thus, they browsed the spaghetti plots of the corresponding time steps in the map view. These spaghetti plots are shown from A to D in Fig.~\ref{fig:temporal_view}, where the complexity of the red lines is disappearing (i.e., few chaos at the subsequent time steps). Thus, the experts concluded that meteorological changes occurred.
	
	\section{Discussion and Limitation}
	
	\par We conducted a semistructured interview with E.1 and E.2 to collect their feedback of our approach.
	
	\par \textbf{System capability.} The experts highly appreciated our framework which significantly boosted their work efficiency. Conventionally, the experts assimilate forecast and satellite data to improve the quality of forecast products. In contrast to the deterministic prediction results, the uncertainty of the ensemble data should be analyzed first for the assimilation of ensemble predictions. Based on the results, they divided the regions or chose different resolutions of satellite data sources~\cite{wang2017multi}. The entire process is time-consuming. With our system, they could easily and intuitively find good results, and the efficiency of region division is greatly improved. Another improvement is that the previous region division is fixed; now, ``\textit{the system provides a quantitative basis of feature distribution and we can divide each time step differently and track the spatial displacement of the uncertainty,}'' said E.2.
	
	\par \textbf{Scalability.} The main bottleneck of our approach is calculating feature distribution. However, our method is based on sampling; thus, we can control the sampling number to achieve a balance between accuracy and computational efficiency. The associated size of the ensemble data we process is determined by the number of grids and ensemble members. We adopt sampling for the number of grids; thus, computational complexity is independent of the number of grids, i.e., denser grids offer us opportunities to obtain refined results. The number of ensemble members is linear with time complexity, thus guaranteeing that complexity growth is maintained within a controllable range. We can also achieve a balance between efficiency and accuracy by adjusting the number of segments of variables. We believe that the sampling number and the number of segments of variables interact with each other, and we plan to explore their relationship in our future work.
	
	\par \textbf{Generalizability.} Although we showcase one weather forecast ensemble data, our approach is applicable to any 2D ensemble data. Furthermore, the proposed variable spreading measure can be extended to 3D data by modifying the design of the 2D map view and the region stability view. For example, for 3D data fitting, we can use \textit{OpenGL} to render the data in a 3D space; for the region stability view, we can borrow and introduce interactive mechanisms~\cite{guo2011wysiwyg} to support interaction with the 3D data. In addition, our framework supports the integration of other kinds of well-established visualization methods. For example, in the 2D map view, we can show other geometric data visualizations.
	
	\par \textbf{Limitation.} We mainly focus on the interaction and selection of spatial location and feature variables. For ensemble forecasting, the change of time-series data also deserves attention. In the case study, we provide users with interactive explorations at various time steps, enabling them to develop insights into the temporal dimension. However, this process is still time consuming and depends heavily on domain expertise and mental memories. In the future, we can integrate the temporal dimension into location and feature for better analysis. Furthermore, our extreme point alignment algorithm is based on the correlation between adjacent time steps; this correlation keeps the local consistency but fails to deliver a remote correlation. For example, if an extreme point disappears in a certain time step and then reappears in a while, the corresponding correlation cannot be detected.

	\section{Conclusion and Future Work}
	\par In this study, we design an interactive visualization framework to help ensemble simulation domain experts explore ensemble data in terms of spatial and variable dimensions. In particular, we propose a measurement method for feature space distribution and design a calculation method based on sampling; this method achieves a balance between computational accuracy and efficiency by controlling the sampling number. Based on the results, we provide different visualization designs to show the feature spreading curve, region stability, and temporal trends. In the future, we plan to extend our calculation method and visualization framework to analyze more than one variable. We also plan to use it for general application scenarios.

	\acknowledgments{
		This research was partially supported by the National Natural Science Foundation of China (Grant Nos. 61972221, 61572274, 61672307), NNW2018-ZT6B12 (National Numerical Windtunnel project), National Key R\&D Program of China (2019YFB1405703) and TC190A4DA/3.
	}
	
		\balance
	
	\bibliographystyle{abbrv}
	\bibliography{template}
\end{document}